\documentclass[journal,10pt]{IEEEtran}
\normalsize

\usepackage{cite}
\usepackage{graphicx}
\usepackage[cmex10]{amsmath}
\usepackage{amssymb}
\usepackage{booktabs}
\usepackage{threeparttable}
\usepackage{algorithm}
\usepackage{algpseudocode}
\usepackage{amsfonts}
\usepackage{xcolor}
\usepackage{color}

\usepackage{tabularx} %

\usepackage{caption}

\usepackage{enumitem}
\usepackage{framed}

\usepackage{dblfloatfix} %

\usepackage{soul}

\usepackage{cancel} %

\usepackage{verbatim}%

\usepackage{makecell}

\usepackage{empheq} %

\usepackage{underoverlap} %

\usepackage{hyperref}
\hypersetup{pdftex,colorlinks=true,allcolors=black}

\usepackage{calc}

\usepackage{textcomp}
\def\BibTeX{{\rm B\kern-.05em{\sc i\kern-.025em b}\kern-.08em
		T\kern-.1667em\lower.7ex\hbox{E}\kern-.125emX}}

\newcommand{\textbb}[1]{{\color{black}#1}}

\newtheorem{remark}{\bf  Remark}

\newtheorem{proposition}{\bf Proposition}
\newtheorem{corollary}{\bf  Corollary}

\newtheorem{lemma}{\bf  Lemma}

\newcommand{\ma}{\mathsf{a}}
\newcommand{\mb}{\mathsf{b}}

\newcommand{\mC}{\mathsf{C}}

\newcommand{\mP}{\mathsf{P}}

\newcommand{\mk}{\mathsf{k}}
\newcommand{\mj}{\mathsf{j}}

\newcommand{\myDef}{\overset{\Delta}{=}}
\newcommand{\myEqualOverset}[1]{\overset{#1}{=}}
\newcommand{\myApprOverset}[1]{\overset{#1}{\approx}}
\newcommand{\myCRLB}[1]{\mathrm{CRLB}\{#1\}}

\newcommand{\imax}{\mathsf{i}_{\mathrm{max}}}
\newcommand{\imin}{\mathsf{i}_{\mathrm{min}}}

\newcommand{\df}{\Delta_f}

\newcommand{\myRound}[1]{\left\lfloor #1 \right\rceil}

\newcommand{\myExp}[1]{\mathbb{E}\left\{#1\right\}}
\newcommand{\myVar}[1]{\mathbb{V}\left\{#1\right\}}

\newcommand{\sincDiscrete}[2]{\mathcal{S}_{#1}\left(#2\right)}

\newcommand{\myRangeLeftNot}[1]{\left(#1\right]}
\newcommand{\myRangeRightNot}[1]{\left[#1\right)}

\newcommand{\myModulo}[2]{\left\langle#1\right\rangle_{#2}}
\newcommand{\myDFT}[2]{\mathcal{Z}_{#1}^{#2}}
\newcommand{\myProb}[1]{\mathbb{P}\left\{ #1 \right\}}
\newcommand{\myCnt}[1]{\mathbb{C}\left\{ #1 \right\}}

\newcommand{\mySpaceTwoMM}{\vspace{2mm}}

\begin{document}

\title{OTFS-Based Joint Communication and Sensing for Future Industrial IoT
}

\author{
	Kai Wu,
	 J. Andrew Zhang,~\IEEEmembership{Senior Member,~IEEE}, %
	Xiaojing Huang,~\IEEEmembership{Senior Member,~IEEE}, and\\
	Y. Jay Guo,~\IEEEmembership{Fellow,~IEEE}
\thanks{This work is partially supported by the Australian Research Council under the Discovery Project Grant DP210101411.}

	\thanks{K. Wu, J. A. Zhang, X. Huang and Y. J. Guo are with the Global Big Data Technologies Centre, University of Technology Sydney, Sydney, NSW 2007, Australia (e-mail: kai.wu@uts.edu.au; andrew.zhang@uts.edu.au; xiaojing.huang@uts.edu.au;   jay.guo@uts.edu.au).}%
\vspace{-1cm}	
}

\maketitle

\begin{abstract}

	Effective wireless communications are increasingly
	important in maintaining the successful closed-loop operation of
	mission-critical industrial Internet-of-Things (IIoT) applications.
	To meet the ever-increasing demands on better wireless communications for IIoT, we propose an 
	orthogonal time-frequency space (OTFS) waveform-based joint
	communication and radio sensing (JCAS) scheme — an energy-efficient
	solution for not only reliable
	communications but also high-accuracy sensing. OTFS has been
	demonstrated to have higher reliability and energy efficiency
	than the currently popular IIoT communication waveforms.
	JCAS has also been highly recommended for IIoT, since it
	saves cost, power and spectrum compared to having two separate radio
	frequency systems. Performing JCAS based on OTFS,
	however, can be hindered by a lack of effective OTFS sensing.
	This paper is dedicated to filling this technology gap. We first design
	a series of echo pre-processing methods that successfully remove the
	impact of communication data symbols in the time-frequency
	domain, where major challenges, like inter-carrier and inter-symbol
	interference and noise amplification, are addressed. Then, we provide a comprehensive analysis of the signal-to-interference-plus-noise ratio (SINR) for sensing and optimize a key parameter of
	the proposed method to maximize the SINR.
	Extensive simulations show that the proposed sensing method
	approaches the maximum likelihood estimator with respect to the estimation accuracy of target parameters and manifests
	applicability to wide ranges of key system parameters. Notably,
	the complexity of the proposed method is only dominated by a
	two-dimensional Fourier transform.

\end{abstract}


\section{Introduction}\label{sec: introduction}
Industrial Internet-of-Things (IIoT), as a key variant of Internet-of-Things, is essentially ``a network of physical objects, systems
platforms and applications that contain embedded technology to communicate and
share intelligence with each other, the external environment and with people'' \cite{chapter_sari2020industrialnowFutureTrends}.
In terms of the way ``things'' communicate in IIoT, wireless links become increasingly preferred over wired counterparts, since the former has some unique advantages, e.g., greater flexibility and higher cost-efficiency \cite{chapter_IIoT_celebi2020wireless}. 
Although some wireless solutions are available already, e.g., ZigBee, LoRa, NB-IoT and LTE-M etc. \cite{IIOT_industrialWirelessNetworkSurvey}, the evolving IIoT constantly raises new challenges. 
A major one is the increasing demand of massive-scale wireless networks for mission-critical IIoT applications, subject to low-latency, high-reliability communications and strict power constraint \cite{IIoT_wirelessNetworkDesign}.

In line with the enormous research effort for enhancing future IIoT, we advocate two techniques --- the joint communication and radar/radio sensing (JCAS) and the orthogonal time-frequency space modulation
 (OTFS) --- both are investigated for 6G recently \cite{Kai_overviewFHMIMO_DFRC2020AES,FanLiu_overview2020TCOM,Kai_rahman2020enablingSurvey,6G_XiaohuYou2021towards,OTFS_Magazine_wei2020orthogonal}. JCAS is proposed to use one hardware platform and one waveform to perform two radio frequency (RF) functions, i.e., communications and sensing, thus saving the cost and power as compared to having two separate transmitters \cite{Kai_overviewFHMIMO_DFRC2020AES}. Moreover, using one waveform for both RF functions can help relieve the increasingly severe spectrum congestion and crowdedness \cite{FanLiu_overview2020TCOM}. 
 Due to its high cost-/power-/spectrum-efficiency,
JCAS has been highly recommended for IoT \cite{JCAS_akan2020internetOfRadar_iot,JCAS_cui2021integratingJCAS4IoT}. 
 In addition, in many IIoT applications, such as intelligent transportation, sensing nodes only need to share some basic parameters, such as ranges and velocities, with other nodes \cite{Kai_rahman2020enablingSurvey}; there is no need for employing sophisticated radar equipments. Therefore, JCAS
 is an attractive and potentially important technology for future IIoT, especially for mission critical ones. 

{To better support the high-mobility use cases in mobile communications, OTFS is proposed to modulate data symbols in the delay-Doppler domain that generally promises longer channel coherence time than the conventional time-frequency domain \cite{OTFS_Magazine_wei2020orthogonal}.
It is worth noting 
that high mobility is becoming a common scenario in IIoT, due to the increasingly use of high-speed platforms, e.g., satellites and drones  \cite{IIoT_droneskumar2021internet}. 
Moreover, OTFS also has a smaller peak to average power ratio (PAPR) than OFDM \cite{OTFS_keskin2021radarTimeDomainICIisi}, which makes OTFS more suitable to work under strict power budget. 
In addition, OTFS can perform better than OFDM and SC-FDMA in IIoT communications that tend to use small-size packets.
This is because, despite packet size, OTFS always spreads information over the whole time-frequency channel in use, thus being able to use the time-frequency-domain diversity to combat channel fading; in contrast, the interleaving and coding, as relied on by OFDM and SC-FDMA to survive channel fading, generally require a large packet size \cite{OTFS_hadani2018otfs_book_chapter}.
Given its advantages over OFDM and SC-FDMA in typical IIoT use scenarios/constraints,
OTFS can potentially be an excellent waveform candidate for future IIoT.}

Since JCAS and OTFS are both promising to further enhance the performance of IIoT communications, combining the two techniques is a natural choice. 
However, one question remains: \textit{how to achieve effective RF sensing using OTFS waveforms with such a low complexity that a power- and computing-limited IIoT device can afford?} Here, OFTS sensing is referred to as estimating the ranges and velocities of targets, e.g., cars and drones etc.

So far, only few published papers dealt with
 OTFS sensing. 
In \cite{OTFS_jcas2020twc}, an iterative optimization procedure is developed to solve the ML-based OTFS sensing problem. In each iteration, 
all the targets are alternatively estimated, one at a time, by searching for the channel matrix that minimizes the signal-to-interference ratio (SIR). The channel matrices are generated based on pre-selected delay-Doppler grids. In \cite{OTFS_yuan2021integratedSensingCOmmunicationOTFS} and \cite{OTFS_Raviteja2019TVT_embeddedPilotChannelEstimation}, matched filtering is performed for OTFS sensing, where the filter coefficients are constructed on pre-defined delay-Doppler grids. Different from the above works \cite{OTFS_jcas2020twc,OTFS_yuan2021integratedSensingCOmmunicationOTFS,OTFS_Raviteja2019TVT_embeddedPilotChannelEstimation} focusing on the delay-Doppler domain, 
the work \cite{OTFS_keskin2021radarTimeDomainICIisi} develops a generalized likelihood ratio test (GLRT) detector in the time domain, where \textbb{the inter-carrier interference (ICI) and inter-symbol interference (ISI) are used to surpass the ambiguity barrier}. 
The GLRT problem is solved by enumerating pre-defined delay-Doppler grids; the ones making the likelihood ratio larger than the threshold are taken as parameter estimations of radar targets. 

\textbb{A common feature of the OTFS sensing methods developed in \cite{OTFS_jcas2020twc,OTFS_yuan2021integratedSensingCOmmunicationOTFS,OTFS_Raviteja2019TVT_embeddedPilotChannelEstimation,OTFS_keskin2021radarTimeDomainICIisi} is that, despite the domain where the estimation problem is formulated,
they estimate target parameters by searching over a set of delay-Doppler grids.
When testing each grid, a metric needs to be calculated with multiplications of large-dimensional matrices/vectors involved\footnote{\textbb{Note that the proposed method also searches over the delay-Doppler domain for target detection. However, we only search for a few peaks over a pre-calculated range-Doppler map; hence, no intensive calculation is required for each searching grid.}}.}
Thus, these methods \cite{OTFS_jcas2020twc,OTFS_yuan2021integratedSensingCOmmunicationOTFS,OTFS_Raviteja2019TVT_embeddedPilotChannelEstimation,OTFS_keskin2021radarTimeDomainICIisi} may not be computationally efficient enough for IIoT devices with strict limits on power and computing resources. 
In addition, the methods in \cite{OTFS_jcas2020twc,OTFS_yuan2021integratedSensingCOmmunicationOTFS,OTFS_Raviteja2019TVT_embeddedPilotChannelEstimation,OTFS_keskin2021radarTimeDomainICIisi} all require a certain number of sensing channel matrices, where the number is equal to that of the overall delay-Doppler grids. Although the channel matrices can be calculated off-line, large storage is then required to save them on-board, which is hardly feasible for IIoT devices.

Different from the existing methods reviewed above, we present a low-complexity OTFS sensing in the time-frequency domain that has not been investigated yet. The idea is inspired by the conventional OFDM sensing, where communication data symbols can be readily removed in the time-frequency domain, then a range-Doppler profile is obtained through a simple two-dimensional Fourier transform and the peaks in the range-Doppler profile are identified for target detection \cite{DFRC_dsss2011procIeee,Kai_ofdmSensingSPM}.  
Despite the close relation between OTFS and OFDM, sensing in the time-frequency domain can be very difficult for OTFS.
{A major challenge is that {communication data symbols} are non-trivial to remove in OTFS sensing.} To improve data rate, OTFS 
tends to use a single cyclic prefix (CP) for a whole block of data symbols \cite{OTFS_Raviteja2019TVTpulseShapingRCP}.
Such a block may equivalently consist of a number of equivalent (in length) OFDM symbols, each with an individual CP. If an OTFS block is processed as a whole, the ICI and ISI issues can be prominent \cite{OTFS_keskin2021radarTimeDomainICIisi}. 
That is, an unknown combination of more than one data symbols can be conveyed by each sub-carrier at the sensing receiver, making the removal of data symbols difficult, if not infeasible. 
{Moreover, even if one manages to recover the sub-carrier orthogonality, the majority of data symbols, conforming to a centered Gaussian distribution in the time-frequency domain, can be near zero. Thus, directly dividing data symbols can largely amplify background noise \cite{DFRC_SC_OFDM}, thus degrading sensing performance.} 
(The distribution of data symbols in the time-frequency domain will be further illustrated in Remark \ref{rmk: gaussian of S[m,n]}.)

In order to realize OTFS-based JCAS for future IIoT, we present a low-complexity OTFS sensing method that has a near-maximum likelihood (ML) performance. Not only the above-mentioned challenges are effectively addressed, but also the optimization of the proposed method is performed to maximize its sensing performance. 
As OTFS communications have been widely studied \cite{OTFS_hadani2018otfs_book_chapter,OTFS_Magazine_wei2020orthogonal}, we only focus on the sensing aspect in this paper. 
Our key contributions include:

\begin{enumerate} [leftmargin=*]
	\item We develop an OTFS sensing framework that is independent of communication data symbols and off-grid. 
	The framework is underpinned by several innovations. \textit{First}, we propose a series of waveform pre-processing methods that recover the sub-carrier orthogonality in the presence of severe ICI and ISI in OTFS. 
	\textit{Second}, we devise an effective way of removing communication data symbols in the time-frequency domain, yet without amplifying the background noise. 
	\textit{Third}, we develop a high-accuracy and off-grid method for estimating ranges and velocities of targets. 	
	
	\item We optimize the key parameter of the proposed OTFS sensing framework to maximize the sensing performance. In doing so, we \textit{first} derive the power expressions 
	of different signal components.  
	\textit{Then} we provide comprehensive analyses, revealing the monotonicity of the power functions of both useful and non-useful signals in terms of the key parameter. 
	\textit{Moreover}, the optimal value of the parameter is derived, showing insights into the impact of system parameters on sensing performance.

	\item We perform extensive simulations to validate the proposed designs and analyses. We \textit{first} provide a set of comparison simulations with the state-of-the-art OTFS sensing method \cite{OTFS_jcas2020twc} that is based on the maximum likelihood (ML) estimation. This first set of results demonstrate the near-ML performance of our OTFS sensing method. Moreover, our design is seen to have an even performance across a wide region of velocity, e.g., $ [0,360] $ kilometer per hour (km/h); whereas, the ML method \cite{OTFS_jcas2020twc} presents a velocity-sensitive estimation performance. We \textit{then} use another set of simulations, validating the precision of the SINR analysis, parameter optimization and the applicability of the proposed method in wide regions of system parameters. 
\end{enumerate}

{The remainder of the paper is organized as follows. Section \ref{sec: signal model} depicts the signal model of OTFS system. Starting with an illustration of the ICI issue in OTFS sensing, Section \ref{sec: sensing framework preprocessing} then proposes a series of pre-processing to recover sub-carrier orthogonality and remove communication information. Section \ref{sec: sensing framework parameter estimation} develops a high-accuracy parameter estimation method and analyzes the overall computational complexity of the proposed sensing scheme. Section \ref{sec: optimizing proposed method tilde M} optimizes the key parameter of the proposed method to maximize the sensing SINR, followed by simulations and conclusions in Sections \ref{sec: simulations} and \ref{sec:conclusions}, respectively.   
}

\section{Signal Model}\label{sec: signal model}

Consider a communication node that transmits a block of $ MN(=M\cdot N) $ data symbols using OTFS modulation. 
Let $ d_{i}~(i=0,1,\cdots,MN-1) $ denote the data symbols which are independently drawn from the same constellation, e.g., 64-QAM. 
These symbols are first placed in a two-dimensional plane, referred to as the delay-Doppler domain. Assume that the delay and Doppler dimensions are discretized into $ N $ and $ M $ grids, respectively. 
Denoting the time duration of $ M $ data symbols as $ T $, the sampling frequency along the Doppler dimension is then $ \frac{1}{T} $, which leads to a Doppler resolution of $ \frac{1}{NT} $. \textbb{Since the time duration of $ M $ grids in the delay dimension is $ T $, the grid resolution along the dimension is $ \frac{T}{M} $.} In OTFS modulation, the data symbols can be mapped from the delay-Doppler domain into the time-frequency domain via the following inverse symplectic Fourier transform \cite{OTFS_hadani2018otfs_book_chapter}, 
\begin{align} \label{eq: S[m,n]}
	S[m,n] & = \frac{1}{\sqrt{MN}} \sum_{k=0}^{N-1}\sum_{l=0}^{M-1} d_{kM+l} e^{\mj2\pi(\frac{nk}{N}-\frac{mT}{M}l\df)}\nonumber\\
	& =\frac{1}{\sqrt{MN}} \sum_{k=0}^{N-1}\sum_{l=0}^{M-1} d_{kM+l} e^{\mj2\pi(\frac{nk}{N}-\frac{ml}{M})},
\end{align}
where $ \mj $ denotes the imaginary unit, $ \df $ is the sub-carrier interval, and the second result is based on the critical sampling, i.e., $ \df T=1 $. 
Modulating $ S[m,n] $ onto the $ m $-th sub-carrier, 
we obtain the following time-domain signal,
\begin{align}\label{eq: s(t)}
	s(t) = \frac{1}{\sqrt{M}}\sum_{n=0}^{N-1}\sum_{m=0}^{M-1} S[m,n] g_{\mathrm{Tx}}(t-nT)e^{\mj 2\pi m\df (t-nT)},
\end{align}
where $ g_{\mathrm{Tx}}(t) $ is a pulse shaping waveform \cite{OTFS_Raviteja2019TVTpulseShapingRCP}. With a CP added to an OTFS block, the transmission waveform of the block becomes
\begin{align} \label{eq: tilde s(t)}
	\tilde{s}(t) = s(\myModulo{t-T_{\mathrm{CP}}}{NT}), ~0\le t\le NT+T_{\mathrm{CP}},
\end{align}
where $ T_{\mathrm{CP}} $ denotes the time duration of a CP, and $ \myModulo{x}{y} $ takes the modulo-$ y $ of $ x $. 

In this paper, we consider the active sensing, where the sensing receiver is co-located with the communication transmitter\footnote{For the transmitter and receiver that are not co-located, extra effort will be required to deal with the estimation ambiguity issues caused by the potential timing and carrier frequency offsets \cite{Kai_rahman2020enablingSurvey}. This is left for future work.}. 
Thus, it is reasonable to assume prefect synchronization and zero frequency offset at the sensing receiver. Consider $ P $ targets. The scattering coefficient, time delay and Doppler frequency of the $ p $-th target are denoted by $ \alpha_p $, $ \tau_p $ and $ \nu_p $, respectively. The target echo consists of the sum of scaled and delayed versions of $ \tilde{s}(t) $, leading to
\begin{align}\label{eq: x(t)}
	x(t) = \sum_{p=0}^{P-1}\alpha_p \tilde{s}(t-\tau_p)e^{\mj 2\pi\nu_p(t-\tau_p)} + w(t), 
\end{align}
where $ w(t) $ is an additive white Gaussian noise. 

\begin{remark}\label{rmk: basic assumptions}
For illustration clarity and analytical tractability, we make some basic assumptions on the signal model. \textit{First}, the single antenna transmitter and receiver is focused on to introduce the new designs and methods. 
\textit{Second}, the rectangular pulse shaping is used.
\textit{Third}, the Swerling-I target model is employed, which indicates that $ \alpha_p~(\forall p) $ is fixed within a coherent processing interval (CPI) of sensing and conforms to a complex Gaussian distribution over CPIs \cite{book_richards2010principlesModernRadar}. Here, we regard the time duration of an OTFS block, i.e., $ NT $, as a CPI. 
\textit{Fourth}, we consider uncorrelated target scattering coefficients, i.e., $ \myExp{\alpha_p\alpha_{p'}}=0,~\forall p\ne p' $, where $ \myExp{\cdot} $ takes expectation. 
\end{remark}

\mySpaceTwoMM

\begin{remark} \label{rmk: gaussian of S[m,n]}
	We remark here on the Gaussian randomness of $ S[m,n] $.  
	With the independently drawn data symbols $ d_i $'s, the $ l $-related summation in (\ref{eq: S[m,n]}), as denoted by $ \tilde{S}[m,k] =  \frac{1}{\sqrt{M}} \sum_{l=0}^{M-1} d_{kM+l} e^{-\mj2\pi\frac{ml}{M}} $, acts as an OFDM modulation and hence conforms to a complex centered Gaussian distribution approximately \cite{Gaussian_OFDMenvelope}. 
	Since the above $ l $-related summation is scaled by $ \frac{1}{\sqrt{M}} $, it acts as an unitary transform. The variance of $ \tilde{S}[m,k] $ then equals to the average power of $ d_i~(\forall i) $, as denoted by $ \sigma_d^2 $. 	
	For any $ m $, we can regard $ \{\tilde{S}[m,0],\tilde{S}[m,1],\cdots,\tilde{S}[m,N-1]\} $	as a white Gaussian noise sequence. 
Since the unitary DFT does not change the Gaussian randomness and its whiteness \cite{book_oppenheim1999discrete}, the $ \frac{1}{\sqrt{N}} $-scaled, $ k $-related summation in (\ref{eq: S[m,n]}) results in a white Gaussian noise sequence; namely, $ S[m,n]\sim\mathcal{CN}(0,\sigma_d^2) $ and is independent over $ \forall m,n $.

\end{remark}
\section{Echo Pre-processing} \label{sec: sensing framework preprocessing}
In this section, we develop a low-complexity sensing method based on OTFS-modulated communication waveform. 
Similar to the transmitter, the critical sampling is performed at the receiver, leading to the following sampling interval 
\[T_{\mathrm{s}}={T}\big/{M}={1}\big/{(M\df)}={1}/{B},\]
where $ B $ denotes the bandwidth. 
The digitally sampled $ x(t) $ with CP removed can be given by
\begin{align} \label{eq: x[i]}
	x[i] &= \sum_{p=0}^{P-1}\alpha_p s\left( \myModulo{i- l_p }{MN}T_{\mathrm{s}} \right)e^{\mj 2\pi\nu_p(i- l_p )T_{\mathrm{s}}} + w(iT_{\mathrm{s}})\nonumber\\
	&= \sum_{p=0}^{P-1}\alpha_p s[\myModulo{i- l_p }{MN}]e^{\mj 2\pi\nu_p(i- l_p )T_{\mathrm{s}}} + w[i],
\end{align}
 where \textbb{$  l_p ={\tau_p/T_{\mathrm{s}}} $}. 
The modulo operator makes $ s[\myModulo{i- l_p }{MN}] $ a circularly shifted version of $ s[i] $ which is the $ i $-th digital sample of $ s(t) $ given in (\ref{eq: s(t)}). 

\textit{If $ MN $ were such a small value that the phase shifts caused by the Doppler frequency remained approximately unchanged across $ i=0,1,\cdots,MN-1 $}, we would be able to take an $ MN $-point DFT of $ x[i] $ given in (\ref{eq: x[i]}), achieving  
\begin{align}\label{eq: X[j]}
	X[j] \approx &\sum_{p=0}^{P-1}\alpha_p e^{-\mj 2\pi\nu_p  l_p  T_{\mathrm{s}}} S[j] e^{-\mj \frac{2\pi j l_p }{MN}}  + W[j],\nonumber\\
	&~~~~~~~~~~~~ ~~~~~j=0,1,\cdots,MN-1
\end{align}
where $ S[j] $ is the $ MN $-point DFT of $ s[i] $ and $ W[j] $ is that of $ w[i] $.
Since $ s[i] $ is known to the sensing receiver, $ S[j] $ is also available. Dividing $ X[j] $ by $ S[j] $ in a point-wise manner, the remaining signal becomes a sum of $ P $ discrete exponential signals, namely $ e^{-\mj \frac{2\pi j l_p }{MN}}~(j=0,1,\cdots,MN-1) $. 
The frequency of any exponential component is solely related to the sample delay of a single target. Thus,
the range estimation is turned into the well studied frequency estimation \cite{Kai_freqEst2020CL,Kai_padeFreqEst2021TVT}. Clearly, being able to obtain (\ref{eq: X[j]}) is helpful and also desirable for estimating target parameters, which, however, is challenging in practice. This is because the value of $ MN $ or $ \nu_p $ can make the variation of the Doppler-induced phase, i.e., $ e^{\mj 2\pi\nu_p(i- l_p )T_{\mathrm{s}}} $ given in (\ref{eq: x[i]}), non-negligible in an OTFS block. The ICI caused by the non-negligible Doppler impact destroys the sub-carrier orthogonality, thus preventing the removal of communication data symbols. 
Below, we introduce some pre-processing on $ x[i] $ to first recover the sub-carrier orthogonality and then remove the data symbols.

\subsection{Reshaping} \label{subsec: reshaping}
In light of the conventional coherent radar signal processing, 
we propose to segment a whole OTFS block (c.f., a radar CPI) into multiple sub-blocks (c.f., radar pulse repetition intervals, PRIs). \textbb{As shown in Fig. \ref{fig: add VCP}, the segmentation is done at the sensing receiver and does not require OTFS transmitter to change its signal format.} 
Let $ \tilde{N} $ denote the number of sub-blocks and $ \tilde{M} $ the number of samples in a sub-block. Note that it is not required to have $ \tilde{M}=M $ or $ \tilde{N}=N $, but $ \tilde{M}\tilde{N}\le MN $ is necessary.
Provided that the following condition holds,
\begin{align} \label{eq: constant doppler phase shift}
	\nu_p\tilde{M}T_{\mathrm{s}}\approx 0,~\forall p,
\end{align}
the target echo in the $ n $-th sub-block can be written into  
\begin{subequations}\label{eq: xn[m]}
	\begin{align}
		x_n[m] & = \sum_{p=0}^{P-1}\tilde{\alpha}_p s\left[ \myModulo{n\tilde{M}+m- l_p }{MN} \right]e^{\mj 2\pi\nu_p(n\tilde{M}+m)T_{\mathrm{s}}} \nonumber\\
		&~~~~+ w_n[m],~\forall n\in[0,\tilde{N}-1],m\in [0,\tilde{M}-1], \label{eq: xn[m] accurate}\\
		& \approx \sum_{p=0}^{P-1}\tilde{\alpha}_p s[n\tilde{M}+m- l_p ]e^{\mj 2\pi\nu_pn\tilde{M}T_{\mathrm{s}}} + w_n[m], \label{eq: xn[m] stop and hop}
	\end{align}
\end{subequations}
where $ \tilde{\alpha}_p=\alpha_pe^{-\mj 2\pi \nu_p  l_p T_{\mathrm{s}}} $ and $ w_n[m] = w[n\tilde{M}+m] $. 
The approximation is obtained by plugging (\ref{eq: constant doppler phase shift}) into (\ref{eq: xn[m] accurate}). Note that the modulo operator, only affecting the first sub-block, is dropped in (\ref{eq: xn[m] stop and hop}) for notational simplicity. 

To help understand the feasibility of the approximation in (\ref{eq: constant doppler phase shift}), we provide a numerical example based on the 5G numerology, where $ \tilde{M}T_{\mathrm{s}}=\frac{1}{480} $ microsecond (ms) \cite{book_ahmadi2019_5G}. Note that $ \tilde{M} $ here resembles the number of samples per symbol. For the carrier frequency of $ 5 $ gigahertz (GHz) and velocity of $ 500 $ kilometer per hour (km/h), the Doppler frequency is $ \nu_p=4.63 $ kilohertz (KHz). Applying these values in (\ref{eq: constant doppler phase shift}), we obtain that $ \nu_p\tilde{M}T_{\mathrm{s}}=9.6\times 10^{-3} \approx 0$. Substituting the result in (\ref{eq: xn[m] accurate}), we further obtain $ e^{\mj 2\pi\nu_p m T_{\mathrm{s}}}\approx 1,~\forall m\in [0,\tilde{M}-1] $.
To reduce the approximation error in (\ref{eq: constant doppler phase shift}), a smaller $ \tilde{M} $ is better. 
As will be seen subsequently, the value of $ \tilde{M} $ also has other impact on the sensing performance of the proposed design. In fact, $ \tilde{M} $ is a critical parameter in our design. In 
Section \ref{sec: optimizing proposed method tilde M}, we will optimize $ \tilde{M} $ to maximize the sensing performance.

\subsection{Virtual Cyclic Prefix} \label{subsec: VCP}

\begin{figure}[!t]
	\centering 
		\includegraphics[width=90mm]{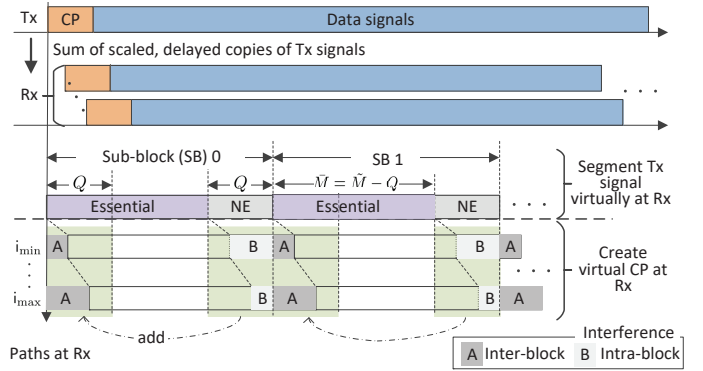}
	\caption{Illustrating the proposed virtual cyclic prefix (VCP), where $ \imin $ denotes the minimum sample delay, $ \imax $ the maximum, and `NE' stands for `non-essential'. \textbb{We emphasize that segmenting and adding VCP are performed on the sensing receiver side, without requiring any changes made on the OTFS transmitter.} 
	}
	\label{fig: add VCP}
\end{figure}

The second design we propose is referred to as the virtual cyclic prefix (VCP). The motivation of this design is to turn the echo signal in each sub-block into a sum of the scaled and circularly shifted versions of the same signal sequence. 
To achieve this, we have seen from (\ref{eq: x[i]}) and (\ref{eq: X[j]}) that a CP is required for the signal sequence (based on which a DFT is taken). 
\textbb{However, the OTFS considered here only has a single CP for a large block; see Fig. \ref{fig: add VCP}. 
Therefore, we propose to create our own CP from the received signals, hence named \textit{virtual} CP. The proposed VCP is performed on 
the sensing receiver side solely and does \textit{not} require the OTFS transmitter to make any changes, e.g., inserting regular CPs as OFDM. 
}

In particular, the proposed design of adding VCP is illustrated in Fig. \ref{fig: add VCP}. Each sub-block is divided into two parts: the first part, consisting of $ (\tilde{M}-Q) $ samples of each transmitted block, is the essential part, while the second part, consisting the remaining $ Q $ samples, is the non-essential part. 
At the sensing receiver, a sub-block is delayed to different extent, after being reflected by different targets. From Fig. \ref{fig: add VCP}, we see that as long as $ Q $ is no smaller than the maximum sample delay, we can add the sampled signals in the non-essential segment onto the first $ Q $ samples of the essential part, and then obtain the circularly shifted versions of the signals in the essential part. Therefore, the non-essential part of each sub-block acts as a VCP. Note that $ Q $ does not have to take the same value as the CP length of the underlying communication system. This is a new flexibility --- \textit{the maximum range detection is not restricted by the CP length of the underlying multi-carrier communication system} --- as ensured by our design and not owned by many previous JCAS schemes\footnote{\textbb{Note that when prominent far scatters exist, the performance of proposed design can be affected, as in the conventional OFDM sensing. However, different from the OFDM sensing restricted to the CP length of the underlying communication system, the VCP length in our design can be flexibly configured to reduce the impact of far scatters (if exist).}}.

\textbb{Note that the practical implementation of the proposed sensing design is similar to the conventional continuous-waveform radar. As mentioned in Section \ref{sec: signal model}, we consider the synchronized transceiver in this paper. Thus, the receiver receives echo signals while the communication transmitter transmits OTFS signals. In practice, the segmentation parameters, i.e., $ \tilde{M} $ and $ Q $, can be pre-determined based on the analysis to be performed in Section \ref{sec: optimizing proposed method tilde M}. So, within the synchronized timing frame, the receiver can segment both the original transmitted signal (as shared by the transmitter) and the echo signal, and can add VCPs for sub-blocks, as described in Fig. \ref{fig: add VCP}.
After signal preprocessing, the receiver can then estimate target parameters using the method to be developed in Section \ref{sec: sensing framework parameter estimation}.  
}

Let $ \tilde{s}_n[m] $ denote the essential signal part of the $ n $-th transmitted sub-block. Referring to Fig. \ref{fig: add VCP}, we have 
\begin{align}\label{eq: tilde sn[m]}
	\tilde{s}_n[m]= s\left[ n\tilde{M}+m \right] ,~ m\in [0,\bar{M}-1]~~(\bar{M}=\tilde{M}-Q),
\end{align}
where $ \bar{M} $ is introduced to simplify notation. 
After adding the proposed VCP for the $ n $-th sub-block, the target echo can be rewritten based on (\ref{eq: xn[m]}), as given by
\begin{align} %
	&\tilde{x}_n[m]  = \sum_{p=0}^{P-1}\tilde{\alpha}_p  \tilde{s}_n\left[ \myModulo{m- l_p }{\bar{M}} \right] e^{\mj 2\pi\nu_pn\tilde{M}T_{\mathrm{s}}} + z^{\mathrm{A}}_n[m] +\nonumber\\
	&~~~~~~~~~~~  z^{\mathrm{B}}_n[m] + w_n[m], ~m=0,1,\cdots,\bar{M}-1,
	\label{eq: tilde xn[m] rewritten}
\end{align}
where $ z^{\mathrm{A}}_n[m] $ denotes the inter-block interference for the $ n $-th sub-block and $ z^{\mathrm{B}}_n[m] $ the intra-block interference. Based on the illustration in Fig. \ref{fig: add VCP}, $ z^{\mathrm{A}}_n[m] $ can be expressed as
\begin{align} \label{eq: zA[n]}
	z^{\mathrm{A}}_n[m] =&  \sum_{p=0}^{P-1}\tilde{\alpha}_p \tilde{s}_{n-1}\left[ m+\tilde{M}- l_p  \right] g_{ l_p }[{m}] e^{\mj 2\pi\nu_p(n-1)\tilde{M}T_{\mathrm{s}}}, \nonumber\\
	&~~~~~~~~~~~~~~~~~~~~~~~~~ m=0,1,\cdots,\imax-1, 
\end{align}
and $ z^{\mathrm{B}}_n[m] $ can be given by 
\begin{align}\label{eq: zB[n]}
	z^{\mathrm{B}}_n[m] = &  \sum_{p=0}^{P-1}\tilde{\alpha}_p \tilde{s}_{n}\left[ m+\bar{M} \right] (1-g_{ l_p }[m]) \times \nonumber\\
	&~~~~~~~ e^{\mj 2\pi\nu_pn\tilde{M}T_{\mathrm{s}}},~m=\imin,\cdots,Q-1, 
\end{align}
where $ g_{x}[{u}]~(x\le Q-1) $ is defined as
\[
g_{x}[{u}]=1,~\text{for }{u}\in [0,x-1];~g_{x}[{u}]=0\text{ for }u\in [x,Q-1].
\]

Note that the two interference terms, $ z^{\mathrm{A}}_n[m] $ and $ z^{\mathrm{B}}_n[m] $, are the price paid for introducing VCP. 
Nevertheless, we notice that $ z^{\mathrm{A}}_n[m] $ and $ z^{\mathrm{B}}_n[m] $ 
can be treated as Gaussian noise uncorrelated with the essential signal part of the $ n $-th sub-block. Lemma \ref{lm: sn[m] zn[m] uncorrelated} provides some useful features of the different signal components in (\ref{eq: tilde xn[m] rewritten}). 
Its proof is provided in Appendix \ref{app: proof of lemma snm and znm uncorrelated}.
In addition, we point out that the impact of the two interference terms can be minimized by properly configuring $ \tilde{M} $. This will be detailed in Section \ref{sec: optimizing proposed method tilde M}.

\vspace{3pt}

\begin{lemma} \label{lm: sn[m] zn[m] uncorrelated}
	\it
	The signal components in (\ref{eq: tilde xn[m] rewritten}), i.e., $ \tilde{s}_n[m] $, $ z^{\mathrm{A}}_n[m] $, $ z^{\mathrm{B}}_n[m] $ and $ w_n[m] $, are approximately independent over $ m=0,1,\cdots,\bar{M}-1 $, and are mutually uncorrelated complex centered Gaussian variables satisfying
	\begin{equation}\label{eq: snm zAnm zBnm wnm Gaussian}
		\begin{gathered} 
			\tilde{s}_n[m]\sim\mathcal{CN}(0,\sigma_d^2) ,~ z^{\mathrm{A}}_n[m]\sim\mathcal{CN}(0,\sigma_{z^{\mathrm{A}}}^2[m]) ,\\
			z^{\mathrm{B}}_n[m]\sim\mathcal{CN}(0,\sigma_{z^{\mathrm{B}}}^2[m]) ,~\mathrm{and}~ w_n[m]\sim\mathcal{CN}(0,\sigma_w^2), 
		\end{gathered}
	\end{equation}
	where $ \sigma_{{d}}^2 $ is the average power of communication data $ d_i~(\forall i) $, $ \sigma_{z^{\mathrm{A}}}^2[m] $ takes the discrete values summarized in Table \ref{tab: value of variance of zAm}, and the possible values of $ \sigma_{z^{\mathrm{B}}}^2[m] $ are given in Table \ref{tab: value of variance of zBm}\footnote{\textbb{Note that the impact of $ Q $ on the interference variances is implicitly shown in the number of different variance values, i.e., the columns of the tables.}}. %
\end{lemma}

\begin{table}[!t]\footnotesize
	\captionof{table}{Values of $ \sigma_{z^{\mathrm{A}}}^2[m] $}
		\vspace{-3mm}
	\begin{center}
		\begin{tabular*}{85mm}{l| llll}
			\hline
			$ m $    & $ \myRangeLeftNot{i_{P-2},i_{P-1}} $ & $ \myRangeLeftNot{i_{P-3},i_{P-2}} $ & $ \cdots $ & $ \myRangeLeftNot{0,i_{0}} $ \\				
			\hline
			$ \sigma_{z^{\mathrm{A}}}^2[m] $    & $ \sigma_d^2 \sigma_{P-1}^2 $ & $ \sigma_d^2 \sum_{p=P-2}^{P-1}\sigma_{p}^2 $ & $ \cdots $ & $ \sigma_d^2 \sum_{p=0}^{P-1}\sigma_{p}^2  $ \\				
			\hline
		\end{tabular*}
		\vspace{-3mm}
	\end{center}
	\label{tab: value of variance of zAm}
\end{table}	
\begin{table}[!t]\footnotesize
	\captionof{table}{Values of $ \sigma_{z^{\mathrm{B}}}^2[m] $}
	\vspace{-3mm}
	\begin{center}
		\begin{tabular*}{75mm}{l| llll}
			\hline
			$ m $    & $ \myRangeRightNot{i_0,i_1} $ & $ \myRangeRightNot{i_{1},i_{2}} $ & $ \cdots $ & $ \myRangeRightNot{i_{P-1},Q-1} $ \\				
			\hline
			$ \sigma_{z^{\mathrm{B}}}^2[m] $    & $ \sigma_d^2 \sigma_{0}^2 $ & $ \sigma_d^2 \sum_{p=0}^{1}\sigma_{p}^2 $ & $ \cdots $ & $ \sigma_d^2 \sum_{p=0}^{P-1}\sigma_{p}^2  $ \\				
			\hline
		\end{tabular*}
		\vspace{-3mm}
	\end{center}
	\label{tab: value of variance of zBm}
\end{table}

\subsection{Removing Communication Information}

Taking the $ \bar{M} $-point DFT of $ \tilde{x}_n[m] $ obtained in (\ref{eq: tilde xn[m] rewritten}) w.r.t. $ m $, we obtain
\begin{align} \label{eq: Xn[l]}
	{X}_n[l] =& \sum_{p=0}^{P-1}\tilde{\alpha}_p\tilde{S}_n[l] e^{-\mj \frac{2\pi l l_p }{\bar{M}}}  e^{\mj 2\pi\nu_pn\tilde{M}T_{\mathrm{s}}}  + Z_n^{\mathrm{A}}[l] + \nonumber\\
	& ~~Z_n^{\mathrm{B}}[l] + W_n[l],~l=0,1,\cdots,\bar{M}-1,
\end{align}
where $ \tilde{S}_n[l] $, $ Z_n^{\mathrm{A}}[l] $, $ Z_n^{\mathrm{B}}[l]$ and $ W_n[l] $ are the $ \bar{M} $-point DFTs (w.r.t. $ m $) of $ \tilde{s}_n[m] $, $ z^{\mathrm{A}}_n[m]$, $z^{\mathrm{B}}_n[m]$ and $w_n[m] $, respectively. 
Since $ \tilde{s}_n[m] $ is known to the sensing receiver, $ \tilde{S}_n[l] $ can be calculated as
\begin{align} \label{eq: tilde Sn[l]}
	\tilde{S}_n[l] = \sum_{m=0}^{\bar{M}-1} \tilde{s}_n[m]\myDFT{\bar{M}}{ml},~~\mathrm{s.t.}~\myDFT{x}{ji}=  e^{-\mj\frac{2\pi ij}{x}}\big/\sqrt{x},
\end{align} 
where $ \myDFT{x}{ji} $ is the $ (j,i) $-th entry of an $ x $-point unitary DFT matrix.
As will be frequently used, the following features of the signal components in (\ref{eq: tilde Xn[l]}) are worth highlighting. The proof is given in Appendix \ref{app: proof of lemma on variances of Snl Znl Wnl}. 

\mySpaceTwoMM

\begin{lemma} \label{lm: variance of Snl ZAnl Zbnl Wnl}
	\it 
	Given $ \forall n,l $, the signal components in (\ref{eq: Xn[l]}) satisfy
	\begin{equation} \label{eq: Snl Wnl Znl Gaussian}
		\begin{gathered} 
			\tilde{S}_n[l]\sim\mathcal{CN}(0,\sigma_{d}^2),~W_n[l]\sim\mathcal{CN}(0,\sigma_{w}^2),\\
			Z_n^{\mathrm{A}}[l] \sim	\mathcal{CN}(0,\sigma_{Z^{\mathrm{A}}}^2),~\mathrm{and}~Z_n^{\mathrm{B}}[l] \sim	\mathcal{CN}(0,\sigma_{Z^{\mathrm{B}}}^2),
		\end{gathered}
	\end{equation}
	where each component is independent over $ \forall l $.
	Assuming that $  l_p ~(\forall p) $ is uniformly distributed in $ [0,Q-1] $, we then have 
	\begin{equation}\label{eq: sigma ZA ^2} 
		\begin{gathered} 
			\sigma_{Z^{\mathrm{A}}}^2=\frac{\imax\sigma_d^2 \sum_{p=0}^{P-1}(p+1)\sigma_p^2}{P\bar{M}},\\
			\sigma_{Z^{\mathrm{B}}}^2 = \frac{(Q-\imin)\sigma_d^2\sum_{p=0}^{P-1}(P-p)\sigma_p^2}{P\bar{M}},
		\end{gathered}
	\end{equation}
	where $ \imax $ and $ \imin $ are the maximum and minimum sample delay, respectively. 
	
\end{lemma}

\mySpaceTwoMM

From (\ref{eq: Xn[l]}), we see that communication information can be suppressed by dividing both sides of the equation by $ \tilde{S}_n[l] $. This, however, can yield significant bursts in the resulted signal, due to the Gaussian randomness of $ \tilde{S}_n[l] $. To reduce the bursts, we propose to use $ \mk \tilde{S}_n[l] $ as the divisor.
The rationale of introducing $ \mk $ is: \textit{multiplying $ \mk(>1) $ can increase $ \myVar{\mk \tilde{S}_n[l]} $ and then decrease $ \myCnt{|\mk \tilde{S}_n[l]|\le 1,~\forall l} $}.
Here, $ \myVar{x} $ denotes the variance of a random variable $ x $ and $ \myCnt{\cdot} $ the event count. 
The following lemma helps configure $ \mk $ and its proof is given in Appendix \ref{app: proof of lemma determining k}.

\mySpaceTwoMM

\begin{lemma}\label{lm: determine k}
	\it 	
	For $ \myProb{|\mk \tilde{S}_n[l]|\le 1} = \epsilon ~(\forall n, l)$, we can set
	\begin{align} \label{eq: k the factor}
		\mk = {1}\Big/\left({\sigma_d \sqrt{\ln\frac{1}{1-\epsilon}}}\right),
	\end{align} 
where $ \epsilon $ denotes a sufficiently small probability. 
\end{lemma}

\mySpaceTwoMM

Dividing both sides of (\ref{eq: Xn[l]}) by $ \mk \tilde{S}_n[l] $, we obtain
\begin{align} \label{eq: tilde Xn[l]}
	& \tilde{X}_n[l] = \mathbb{I}_n[l]\left( \sum_{p=0}^{P-1}\frac{\tilde{\alpha}_p}{\mk} e^{-\mj \frac{2\pi l l_p }{\bar{M}}}  e^{\mj 2\pi\nu_pn\tilde{M}T_{\mathrm{s}}}  + \tilde{Z}_n^{\mathrm{A}}[l] + \right.\nonumber\\
	& \left. \textcolor{white}{\sum_{p=0}^{P}}\tilde{Z}_n^{\mathrm{B}}[l] + \tilde{W}_n[l]\right),~\mathbb{I}_n[l]=\left\{ \begin{array}{ll}
		0&\text{ if }|\mk \tilde{S}_n[l]|\le 1\\
		1&	\text{ otherwise }
	\end{array}
 \right.,
\end{align}
where
the intermediate variables are given by
\begin{equation}\label{eq: breve ZnAl ZnBl Wnl}
	\begin{gathered}
		\tilde{Z}_n^{\mathrm{A}}[l]=Z^{\mathrm{A}}_n[l]/(\mk \tilde{S}_n[l]),~\tilde{Z}_n^{\mathrm{B}}[l]=Z^{\mathrm{B}}_n[l]/(\mk \tilde{S}_n[l]),\\
		\tilde{W}_n[l] = W_n[l]/(\mk \tilde{S}_n[l]).
	\end{gathered}
\end{equation}
In case $ |\mk \tilde{S}_n[l]|< 1 $ (although unlikely given a properly selected $ \mk $), $ \mathbb{I}_n[l] $ can help prevent noise enhancement. 

\textbb{We remark that the way we remove communication information is similar to the widely used OFDM sensing \cite{DFRC_dsss2011procIeee}. However, we emphasize that it is our design proposed in Sections \ref{subsec: reshaping} and \ref{subsec: VCP} that enable OTFS sensing to be performed in such a simple manner. On the other hand, we notice that 
when dividing communication data symbols in OFDM sensing, the noise enhancement is not an issue in general, as PSK (with constant modulus) is assumed in most OFDM sensing works. In contrast, we are dealing with OTFS modulation that obtains the signals in the time-frequency domain through a symplectic Fourier transform; see (\ref{eq: S[m,n]}), making noise enhancement issue rather severe. But thanks to the design and analysis in this section, the issue can be greatly relieved.  
}

\textbb{We also remark that the pointwise division (PWD) performed in (\ref{eq: tilde Xn[l]}) can be replaced by pointwise product (PWP). As PWP is performed in the frequency domain, taking the IDFT of the PWP result leads to the cyclic cross-correlation of the corresponding time-domain signals, similar to the matched filtering in conventional radar processing. 
The SINRs in the range-Doppler maps (RDM) generated by PWD and PWP	have interesting relations, as revealed in \cite{Kai_integrateSensingIntoCom2021JSAC}. In particular, in low SNR regions, the PWP-SINR is higher than PWD-SINR, while the relation reverses in high SNR regions. Interested readers are referred to \cite{Kai_integrateSensingIntoCom2021JSAC} for more details. 
}

\section{Target Parameter Estimation} \label{sec: sensing framework parameter estimation}

Enabled by the proposed pre-processing on the target echo, the resulted signal $ \tilde{X}_n[l] $ presents a clear structure. In particular, we see from (\ref{eq: tilde Xn[l]}) that $ e^{-\mj \frac{2\pi l l_p }{\bar{M}}} $ and $ e^{\mj 2\pi\nu_pn\tilde{M}T_{\mathrm{s}}} $ are two single-tone signals whose frequencies are related to target range and velocity, respectively. Therefore,  parameter estimation in OTFS sensing is turned into estimating the center frequencies of the multi-tone signal $ \tilde{X}_n[l] $. To fulfill the task, we develop a low-complexity and high-accuracy method, combining the commonly used estimate-and-subtract strategy \cite{FreqEst_AMexMultipleTones2017SP,FreqEst_multiToneQSE2020} with the frequency estimator we recently proposed in \cite{Kai_padeFreqEst2021TVT} for a single-tone signal. Below, we first illustrate the parameter estimation method and then analyze the computational complexity of the whole OTFS sensing scheme. 

\subsection{Parameter Estimation Method}

The overall estimation procedure is
 summarized in Algorithm \ref{alg: over target parameter estimation}\footnote{As commonly assumed in the work centered on parameter estimation \cite{FreqEst_AMexMultipleTones2017SP,FreqEst_multiToneQSE2020}, the total number of targets, i.e., $ P $, is taken as a known input.
	In practice, $ P $ can be estimated through well-developed techniques like the Akaike information criterion (AIC) and the minimum description length (MDL) \cite{book_van2004optimum}. However, we remark that a
low-complexity detection method that is suitable for IIoT devices is worth investigating.
}. 
{In each iteration}, 
we always estimate the parameters of the presently strongest target, as done in Steps 3) to 6) of the table; reconstruct the echo signal of the target and subtract the reconstructed echo signal, as done in Step 2); and re-perform the above three steps for the next strongest target. For ease of illustration, we assume that the $ p $-th $ (\forall p) $ target has stronger echo than the $ p' $-th $ (\forall p'> p) $ target, namely $ |\tilde{\alpha}_0|>|\tilde{\alpha}_1|>\cdots>|\tilde{\alpha}_{P-1}| $. Thus, the signal used for estimating the $ p $-th target can be given by 
\begin{align} \label{eq: tilde X_n^(p)[l]}
	&\tilde{X}_n^{(p)}[l] = \tilde{X}_n^{(p-1)}[l] - 
	\frac{\hat{\alpha}_{p-1}}{\mk} e^{-\mj \frac{2\pi l \hat{l}_{p-1} }{\bar{M}}}  e^{\mj 2\pi\hat{\nu}_{p-1} n\tilde{M}T_{\mathrm{s}}}\nonumber\\
	\mathrm{s.t.}~&~ p\ge 1,~\tilde{X}_n^{(0)}[l] =  \tilde{X}_n[l]\text{ given in (\ref{eq: tilde Xn[l]})},
\end{align}
where $ \tilde{X}_n^{(p-1)}[l] $ denotes the signal used for estimating the $ (p-1) $-th target, and $ \hat{\alpha}_{p-1},\hat{l}_{p-1} $ and $ \hat{\nu}_{p-1} $ denote the estimated parameters of the target.

\begin{table}[!t]\small
	\captionof{algorithm}{\small Overall Parameter Estimation}
	\vspace{-3mm}
	\begin{center}%
		\begin{tabular}{p{8.5cm}}
			\hline

			Input: $ \tilde{X}_n[l] $, as given in (\ref{eq: tilde Xn[l]}), and $ P $ (the number of targets)
			
			\begin{enumerate}[leftmargin=*]
				
				\item 
				Initialize $ p=0  $;
				
				\item Update $ \tilde{X}_n^{(p)}[l] $ as done in (\ref{eq: tilde X_n^(p)[l]});
				
				\item Estimate $(\tilde{l}_p,\tilde{n}_p)  $ as done in (\ref{eq: tilde l_p, tilde n_p: max});

				\item Estimate $ \delta_p $ and $ \epsilon_p $ by running Algorithm \ref{alg: estimating delta_p};
				
				\item Obtain the estimates of $ l_p $ and $ \nu_p $ as given in (\ref{eq: hat l_p hat nu_p});
				
				\item Estimate $ \alpha_p $ as done in (\ref{eq: hat alpha_p});
				
				\item End if $ p=P-1 $; otherwise, $ p=p+1 $ and go to Step 2).
				
				\vspace{-3mm}

			\end{enumerate}

				\\
				\hline
			
		\end{tabular}
		\vspace{-5mm}	
	\end{center}
	\label{alg: over target parameter estimation}
\end{table}

As mentioned earlier, $ e^{-\mj \frac{2\pi l l_p }{\bar{M}}} $ and $ e^{\mj 2\pi\nu_pn\tilde{M}T_{\mathrm{s}}} $ in (\ref{eq: tilde Xn[l]}) are exponential signals along $ l $ and $ n $, respectively. Therefore, we can identify their center frequencies through a two-dimensional DFT:
\begin{align} \label{eq: breve X_{tilde n}^{(p)} (tilde l)}
	& \breve{X}_{\tilde{n}}^{(p)}[\tilde{l}] = \sum_{n=0}^{\tilde{N}-1}\sum_{l=0}^{\bar{M}-1} \tilde{X}_n^{(p)}[l]\myDFT{\bar{M}}{-l\tilde{l}}\myDFT{\tilde{N}}{n\tilde{n}} \\
	& = \frac{\tilde{\alpha}_p}{\mk} \sincDiscrete{\bar{M}}{\tilde{l}- l_p } \sincDiscrete{\bar{N}}{\underbrace{\tilde{N}\tilde{M}T_{\mathrm{s}}\nu_p}_{n_p}-\tilde{n}} +  \Xi_{\tilde{n}}^{(p)}[\tilde{l}], \nonumber
\end{align}
where $ \myDFT{\bar{M}}{-l\tilde{l}}$ and $\myDFT{\tilde{N}}{n\tilde{n}} $ are the DFT bases defined in (\ref{eq: tilde Sn[l]}). Note that all irrelevant terms, including interference plus noise terms given in (\ref{eq: tilde Xn[l]}) and the echoes of weaker targets $ p'~(=p+1,\cdots,P-1) $, are absorbed in $ \Xi_{\tilde{n}}^{(p)}[\tilde{l}] $ for brevity. Moreover, the function $ \sincDiscrete{x}{y} $ in (\ref{eq: breve X_{tilde n}^{(p)} (tilde l)}) is defined as
\begin{align} \label{eq: S_x(y)}
	\sincDiscrete{x}{y} = \frac{1}{\sqrt{x}} \frac{\sin\left( \frac{x}{2}\frac{2\pi y}{x} \right)}{\sin\left( \frac{1}{2}\frac{2\pi y}{x} \right)}e^{\mj \frac{x-1}{2} \frac{2\pi y}{x}}.
\end{align} 
Note that $ \sincDiscrete{x}{y} $ is a discrete sinc function which is maximized at $ y=0 $. Therefore, the integer parts of $ (l_p , n_p) $ can be estimated by identifying the maximum of $ \left|\breve{X}_{\tilde{n}}^{(p)}[\tilde{l}]\right|^2 $, i.e.,
\begin{align} \label{eq: tilde l_p, tilde n_p: max}
	(\tilde{l}_p,\tilde{n}_p) :~\max{}_{\forall \tilde{l},\tilde{n}} \left|\breve{X}_{\tilde{n}}^{(p)}[\tilde{l}]\right|^2.
\end{align}
To obtain high-accuracy estimations of target parameters, the fractional parts of $ (l_p , n_p)~\forall p $, as denoted by $ (\delta_p, \epsilon_p) $, also need to be estimated. 
With reference to \cite{Kai_padeFreqEst2021TVT}, we develop below the methods for estimating $ \delta_p$ and $ \epsilon_p $.

\begin{table}[!t]\small
	\captionof{algorithm}{\small  Estimating $ \delta_p $ (or $ \epsilon_p{}^{\dagger} $)}
	\vspace{-3mm}
	\begin{center}%
		\begin{tabular}{p{8.5cm}}
			\hline
			\begin{enumerate}[leftmargin=*]\renewcommand{\labelenumi}{{\arabic{enumi})}}
				\vspace{-2mm}
				
				\item Input: $ \tilde{X}_n^{(p)}[l] $ given in (\ref{eq: tilde X_n^(p)[l]}), $ (\tilde{l}_p,\tilde{n}_p) $ in (\ref{eq: tilde l_p, tilde n_p: max}) and $ N_{\mathrm{iter}} $ (the maximum number of iterations);
				
				\item Initialize: $ \hat{\delta}_p^{(0)}=0.25d_p $ and $ i=1 $, where $ d_p $ is given in (\ref{eq: d_p sign test});
				
				\item Interpolate the DFT coefficients at $ \tilde{l}=\tilde{l}_{p}^{\pm}\myDef \tilde{l}_p+\hat{\delta}_p^{(i-1)} \pm 0.25 $, leading to $ \breve{X}_{\tilde{n}_p}^{(p)}[\tilde{l}_{p}^{\pm}]  $ given in (\ref{eq: breve X _{tilde n_p} [tilde l_p]});
				
				\item Construct the ratio of $ \rho_p^{(i)}=\frac{|x_+|^2 - |x_-|^2}{|x_+|^2 + |x_-|^2},~\mathrm{s.t.}~x_{\pm}\myDef \breve{X}_{\tilde{n}_p}^{(p)}[\tilde{l}_{p}^{\pm}] $;
				
				\item Calculate $ r_i~(i=0,1,2) $ as done in (\ref{eq: r0 r1 r2 roots of cubic equation});
				
				\item Estimate $ \xi_p^{(i)} $ as $ \hat{\xi}_p^{(i)} = r_{i^*},~\mathrm{s.t.}~i^*=\mathrm{argmin}{}_{i=0,1,2}~|r_i| $;	
				
				\item Update $ \hat{\delta}_p^{(i)}=\hat{\delta}_p^{(i-1)} + \xi_p^{(i)} $; 
				
				\item Set $ i=i+1 $ and go back to Step 3), if $ i<N_{\mathrm{iter}} $;
				
				\item Output: the final estimate $ \hat{\delta}_p = \hat{\delta}_p^{(i-1)}+\hat{\xi}_p^{(i)} $.
				
				\vspace{-2mm}
				
			\end{enumerate}\\
			\hline
			$ {}^{\dagger} $	{\footnotesize When estimating $ \epsilon_p $, $ \delta_p $ above is replaced by $ \epsilon_p $, $ \xi_p^{(i)} $ by $ \eta_p^{(i)} $, $ {\hat{\delta}}_p^{(i)} $ by $ {\hat{\epsilon}}_p^{(i)} $; see (\ref{eq: eta_p^(i)}). Moreover, $ \breve{X}_{\tilde{n}_p}^{(p)}[\tilde{l}_{p}^{\pm}] $ in Steps 3) and 4) becomes $ \breve{X}_{\tilde{n}_p^{\pm}}^{(p)}[\tilde{l}_{p}] $ given in (\ref{eq: breve X _{tilde n_p +-} [tilde l_p]}). The remaining steps can be run without changes. However, note that $ c_1 $, $ c_3 $ and $ c_5 $ required in (\ref{eq: r0 r1 r2 roots of cubic equation}) now become the Taylor coefficients of the function given in (\ref{eq: f(eta_p(i))}). 	}		
			\\
			\hline
			
		\end{tabular}
		\vspace{-5mm}	
	\end{center}
	\label{alg: estimating delta_p}
\end{table}	

We start with $ \delta_p$. As summarized in Algorithm \ref{alg: estimating delta_p}, its estimation is performed iteratively. 
At iteration $ i(\ge 1) $, we have the estimate of $ \delta_p $ from the iteration $ (i-1) $, as denoted by $ \hat{\delta}_p^{(i-1)} $. Let $ \xi_p^{(i)}=\delta_p - \hat{\delta}_p^{(i-1)} $ denote the estimation error which is estimated in Steps 3) to 6). 
In Step 3), the interpolated DFT coefficients are given by
\begin{align} \label{eq: breve X _{tilde n_p} [tilde l_p]}
	& \breve{X}_{\tilde{n}_p}^{(p)}[\tilde{l}_{p}^{\pm}] = \sum_{n=0}^{\tilde{N}-1}\sum_{l=0}^{\bar{M}-1} \tilde{X}_n^{(p)}[l]\myDFT{\bar{M}}{-l\tilde{l}_p^{\pm}}\myDFT{\tilde{N}}{n\tilde{n}_p} =    \Xi_{\tilde{n}_p}^{(p)}[\tilde{l}_{p}^{\pm}] \nonumber\\
	&~~~~~~~~~~~~~~~~~~~~~~~~ \frac{\tilde{\alpha}_p}{\mk} \sincDiscrete{\bar{M}}{\tilde{l}_{p}^{\pm} - l_p } \sincDiscrete{\bar{N}}{n_p-\tilde{n}_p},
\end{align}
where $ (\tilde{l}_p,\tilde{n}_p) $ is obtained in (\ref{eq: tilde l_p, tilde n_p: max}) and $ \tilde{l}_{p}^{\pm}\myDef \tilde{l}_p+\hat{\delta}_p^{(i-1)} \pm 0.25 $.
 In Step 4), the ratio $ \rho_p^{(i)} $ is constructed such that it can be regarded as a noisy value of the function $ f(\xi_p^{(i)}) $ given in (\ref{eq: f(xi_p^{(i)})}). Note that the result ``$ \myEqualOverset{(a)} $'' in (\ref{eq: f(xi_p^{(i)})})  is obtained by suppressing the common terms of the numerator and denominator and by replacing $ \tilde{l}_{p}^{\pm}-l_p $ with 
 \[
 \tilde{l}_{p}^{\pm}-l_p = \tilde{l}_p+\hat{\delta}_p^{(i-1)} \pm 0.25 - (\tilde{l}_p+\delta_p) = -\xi_p^{(i)} \pm 0.25.
 \]
Equating $ f(\xi_p^{(i)})$ and $\rho_p^{(i)} $, $ \xi_p^{(i)} $ can be estimated by solving the equation. As derived in \cite{Kai_padeFreqEst2021TVT}, there are 
three roots,  
 \begin{figure*}[!t]
 	\begin{align} \label{eq: f(xi_p^{(i)})}
		 	f(\xi_p^{(i)}) = \frac{
		 		\begin{array}{l}
		 			\left|  \frac{\tilde{\alpha}_p}{\mk} \sincDiscrete{\bar{M}}{\tilde{l}_{p}^{+} - l_p } \sincDiscrete{\bar{N}}{n_p-\tilde{n}_p} \right|^2 \\
		 			~~~~~~~~~~~~~~~~~~~~~ - \left|  \frac{\tilde{\alpha}_p}{\mk} \sincDiscrete{\bar{M}}{\tilde{l}_{p}^{-} - l_p } \sincDiscrete{\bar{N}}{n_p-\tilde{n}_p} \right|^2
		 		\end{array}
		 	}{
		\begin{array}{l}
			\left|  \frac{\tilde{\alpha}_p}{\mk} \sincDiscrete{\bar{M}}{\tilde{l}_{p}^{+} - l_p } \sincDiscrete{\bar{N}}{n_p-\tilde{n}_p} \right|^2 \\
			~~~~~~~~~~~~~~~~~~~~~ + 
			\left|  \frac{\tilde{\alpha}_p}{\mk} \sincDiscrete{\bar{M}}{\tilde{l}_{p}^{-} - l_p } \sincDiscrete{\bar{N}}{n_p-\tilde{n}_p} \right|^2 
		\end{array}
	} 
	\myEqualOverset{(a)} \frac{
		\left|  \sincDiscrete{\bar{M}}{\xi_p^{(i)} -0.25 }  \right|^2 - \left|   \sincDiscrete{\bar{M}}{ \xi_p^{(i)} + 0.25 }  \right|^2 } 
	{
	\left|   \sincDiscrete{\bar{M}}{ \xi_p^{(i)} - 0.25 }  \right|^2 + 
	\left|   \sincDiscrete{\bar{M}}{ \xi_p^{(i)} + 0.25 }  \right|^2 
}
 	\end{align}
 \vspace{-5mm}
 \end{figure*}
\begin{align}\label{eq: r0 r1 r2 roots of cubic equation}
	&~r_1 = -k_2/3 + 2B, ~r_2 = -k_2/3 - B + D, \nonumber\\
	&~r_3 = -k_2/3 - B - D, \nonumber\\
	\mathrm{s.t.}~&~B=(S+T)/2,~D={\sqrt{3}}(S-T)\mj/2,\nonumber\\
	&~ S=\sqrt[3]{R+\sqrt{D}},~T=\sqrt[3]{R-\sqrt{D}},~D=Q^3 + R^2\nonumber\\
	&~R={(9k_1k_2-27k_0-2k_2^3)}/{54},~Q={(3k_1-k_2^2)}/{9},\nonumber\\
	&~ k_2=-{\rho b_2}/{a_3},k_1={a_1}/{a_3},k_0=- {\rho}/{a_3},\nonumber\\
	&~a_1 = c_1,a_3=c_3-{c_1c_5}\big/{c_3},b_2={c_5}\big/{c_3},
\end{align}
where $ c_1 $, $ c_3 $ and $ c_5 $ are the coefficients of the first, third and fifth power terms in the Taylor series of $ f(\xi_p^{(i)}) $ at $ \xi_p^{(i)}=0 $. 
Only one root provides the estimate of $ \xi_p^{(i)} $, as determined in Step 6). Next, we update the estimate of $ \delta_p $ in Step 7), check the stop criterion in Step 8) and, if unsatisfied, run another iteration. 
Above is the general iteration procedure. As for the initialization of the algorithm, $ d_p $ is the result of the following sign test:
\begin{align} \label{eq: d_p sign test}
	d_p=\mathrm{sign}\left\{ \left[\breve{X}_{\tilde{n}_p}^{(p)}[\tilde{l}_p-1]-\breve{X}_{\tilde{n}_p}^{(p)}[\tilde{l}_p+1]\right] \left(\breve{X}_{\tilde{n}_p}^{(p)}[\tilde{l}_p]\right)^*\right\}, 
\end{align}
where $ \breve{X}_{\tilde{n}_p}^{(p)}[\tilde{l}_p] $ is obtained by plugging $ \tilde{n}=\tilde{n}_p $ and $ \tilde{l} = \tilde{l}_p $ into (\ref{eq: breve X_{tilde n}^{(p)} (tilde l)}) and likewise $ \breve{X}_{\tilde{n}_p}^{(p)}[\tilde{l}_p\pm 1] $ is obtained. Above, the estimation steps are given with the rationales suppressed for brevity. Interested readers may refer to \cite{Kai_padeFreqEst2021TVT} for more details.

We notice that $ \epsilon_p $ can also be estimated as done in Algorithm \ref{alg: estimating delta_p}, with necessary changes pointed out in the table. For $ \epsilon_p $, the estimation error in the iteration $ i(\ge 1) $ is denoted by
\begin{align}\label{eq: eta_p^(i)}
	\eta_p^{(i)} = \epsilon_p-\hat{\epsilon}_p^{(i-1)}, 
\end{align} 
where $ \hat{\epsilon}_p^{(i-1)} $ is the estimate from the previous iteration. 
The DFT interpolation in Step 3) now happens at $ \tilde{n}=\tilde{n}_p^{\pm}\myDef\tilde{n}_p+\hat{\epsilon}_p^{(i-1)} \pm 0.25  $ and the interpolated coefficients are given by 
\begin{align} \label{eq: breve X _{tilde n_p +-} [tilde l_p]}
	& \breve{X}_{\tilde{n}_p^{\pm}}^{(p)}[\tilde{l}_{p}] = \sum_{n=0}^{\tilde{N}-1}\sum_{l=0}^{\bar{M}-1} \tilde{X}_n^{(p)}[l]\myDFT{\bar{M}}{-l\tilde{l}_p}\myDFT{\tilde{N}}{n\tilde{n}_p^{\pm}} =    \Xi_{\tilde{n}_p^{\pm}}^{(p)}[\tilde{l}_{p}] \nonumber\\
	&~~~~~~~~~~~~~~~~~~~~~~~~ \frac{\tilde{\alpha}_p}{\mk} \sincDiscrete{\bar{M}}{\tilde{l}_{p} - l_p } \sincDiscrete{\bar{N}}{n_p-\tilde{n}_p^{\pm}},
\end{align}
which can be likewise interpreted as $ \breve{X}_{\tilde{n}_p}^{(p)}[\tilde{l}_{p}^{\pm}] $ given in (\ref{eq: breve X _{tilde n_p} [tilde l_p]}). 
Moreover, corresponding to $ f(\xi_p^{(i)}) $ given in (\ref{eq: f(xi_p^{(i)})}), we now have 
\begin{align} \label{eq: f(eta_p(i))}
	f(\eta_p^{(i)}) = \frac{
		\left|  \sincDiscrete{\bar{N}}{\eta_p^{(i)} -0.25 }  \right|^2 - \left|   \sincDiscrete{\bar{N}}{ \eta_p^{(i)} + 0.25 }  \right|^2 } 
	{
		\left|   \sincDiscrete{\bar{N}}{ \eta_p^{(i)} - 0.25 }  \right|^2 + 
		\left|   \sincDiscrete{\bar{N}}{ \eta_p^{(i)} + 0.25 }  \right|^2 
	}
\end{align}
which is used for computing the three roots according to (\ref{eq: r0 r1 r2 roots of cubic equation}).

After running Algorithm \ref{alg: estimating delta_p}, we obtain $ \hat{\delta}_p $ and $ \hat{\epsilon}_p $. Adding them with $ \tilde{l}_p $ and $ \tilde{n}_p $ given in (\ref{eq: tilde l_p, tilde n_p: max}), the estimates of $ {l}_p $ and $ {\nu}_p $ can be written as
\begin{align} \label{eq: hat l_p hat nu_p}
	\hat{l}_p = \tilde{l}_p + \hat{\delta}_p,~\hat{\nu}_p = (\tilde{n}_p + \hat{\epsilon}_p)\Big/(\tilde{N}\tilde{M}T_{\mathrm{s}}), 
\end{align}
where the relation between $ n_p $ and $ \nu_p $ is shown in (\ref{eq: breve X_{tilde n}^{(p)} (tilde l)}). 
Substituting $ \hat{l}_p $ and $ \hat{\nu}_p $ into (\ref{eq: breve X_{tilde n}^{(p)} (tilde l)}), 
the complex coefficient $ \tilde{\alpha}_p $ can be estimated as
\begin{align}\label{eq: hat alpha_p}
	\hat{\alpha}_p = \mk \breve{X}_{\tilde{n}_p}^{(p)}[\tilde{l}_p]\Big/\left( \sincDiscrete{\bar{M}}{\tilde{l}_p - \hat{l}_p } \sincDiscrete{\bar{N}}{\hat{n}_p - \tilde{n}_p } \right),
\end{align}
where the values of the two discrete sinc functions can be readily calculated based on (\ref{eq: S_x(y)}).

\subsection{Computational Complexity}

Next, we analyze the computational complexity (CC) of the proposed OTFS and compare it with the existing methods. For pre-processing the echo, as illustrated in Section \ref{sec: sensing framework preprocessing}, the only computation-intensive computations are given in (\ref{eq: Xn[l]}) and (\ref{eq: tilde Xn[l]}) which perform $ \tilde{N} $ numbers of $ \bar{M} $-dimensional DFT and an $ \tilde{N}\bar{M} $-size point-wise division, respectively. Thus, the CC of echo pre-processing is in the order of 
$ \mathcal{O}\{ \tilde{N}\bar{M}\log\bar{M} \} $.
For the proposed parameter estimation method, as summarized in Algorithm \ref{alg: over target parameter estimation}, its CC is analyzed below.
\begin{itemize}[leftmargin=*]
	\item Step 3) involves a computation of a two-dimensional DFT, as done in (\ref{eq: breve X_{tilde n}^{(p)} (tilde l)}), incurring the CC of $ \mathcal{O}\{\tilde{N}\bar{M}\log(\bar{M}\tilde{N})\} $.
Step 3) also requires $ P $ times of searching for the peak in an $ \tilde{N}\times \bar{M} $-dimensional range-Doppler profile, yielding the CC of $ \mathcal{O}\{P\tilde{N}\log\bar{M} \} $. Since $ \tilde{N}\bar{M}\log(\bar{M}\tilde{N})\gg P\tilde{N}\log\bar{M} $, the overall CC of Step 3) in Algorithm \ref{alg: over target parameter estimation} is $ \mathcal{O}\{\tilde{N}\bar{M}\log(\bar{M}\tilde{N})\} $; 
	
	\item Step 4) runs Algorithm \ref{alg: estimating delta_p} for $ P $ times, one for each targets. In each time, the algorithm is performed twice, one for range estimation and another for Doppler. Moreover, the CC of Algorithm \ref{alg: estimating delta_p} is dominated by the interpolation in Step 3) \cite{Kai_padeFreqEst2021TVT} and is given by $ \mathcal{O}\{ (2N_{\mathrm{iter}}+1)\bar{M} \} $ and $ \mathcal{O}\{ (2N_{\mathrm{iter}}+1)\tilde{N} \} $, respectively, for range and Doppler estimations. Thus, Step 4) of Algorithm \ref{alg: over target parameter estimation} has a CC of $ \mathcal{O}\{(2N_{\mathrm{iter}}+1)P x \} $ with $ x=\max\{\bar{M},\tilde{N}\} $.
\end{itemize}
As will be shown in Section \ref{sec: simulations}, a small value of $ N_{\mathrm{iter}} $, e.g., five, can ensure a near-ML estimation performance of Algorithm \ref{alg: estimating delta_p}. Moreover, $ P $ is typically smaller than $ \tilde{N} $ or $ \bar{M} $ in practice. In summary, we can assert that the CC of the proposed OTFS sensing scheme is dominated by that of the two-dimensional DFT performed in Step 3) of Algorithm \ref{alg: over target parameter estimation}.

Since the ML-based OTFS sensing method \cite{OTFS_jcas2020twc} will be employed as a benchmark in the simulations, we provide below its CC as a comparison. As a lower bound on its CC, we only take into account the first iteration of the ML method under a single target assumption. From (22) and its context in \cite{OTFS_jcas2020twc}, we can see that the number of range-Doppler grids is $ N_{\mathrm{CP}}N $, where $ N_{\mathrm{CP}} $ is the number of samples in the CP of the underlying OTFS communication system and $ N $ is the size of the Doppler dimension. For each grid, the likelihood ratio given in \cite[(21)]{OTFS_jcas2020twc} needs to be calculated, yielding the CC of $ \mathcal{O}\{(MN)^2\} $. Thus, a lower bound of the overall CC of the ML method can be given by $ \mathcal{O}\{N_{\mathrm{CP}}M^2N^3\} $. Based on (\ref{eq: tilde sn[m]}) and the text above (\ref{eq: constant doppler phase shift}), we have $ \bar{M}=\tilde{M}-Q<\tilde{M} $ and $ \tilde{N}\bar{M}<\tilde{N}\bar{M}<MN $, and hence $ \tilde{N}\bar{M}\log(\bar{M}\tilde{N})\ll N_{\mathrm{CP}}M^2N^3 $. 
This demonstrates the high computational efficiency of the proposed OTFS sensing.

\textbb{Before ending the section, we remark on the potential extension of the proposed method to a multiple-input and multiple-output (MIMO) system with independent signals transmitted from all antennas. Based on Remark \ref{rmk: gaussian of S[m,n]}, we know that the transmitted signals have low mutual cross-correlations. This suggests that we can perform the methods developed in Sections \ref{sec: sensing framework preprocessing} and \ref{sec: sensing framework parameter estimation} on 
the signal received by each receiver antenna using each transmitted signal as (\ref{eq: tilde s(t)}), as in a conventional orthogonal MIMO radar. However, $ \tilde{x}_n[m] $ obtained in (\ref{eq: tilde xn[m] rewritten}) will have an extra interference term, owing to the cross-correlations among transmitted signals. Though the interference does not affect the way the proposed methods are performed, it can degrade sensing performance. Thus, optimizing the transmitted waveforms to reduce their cross-correlations, subject to a satisfactory communication performance, can be an interesting future work. 
}

\section{Optimizing Proposed Estimation Method}
\label{sec: optimizing proposed method tilde M}

In this section, we optimize the proposed sensing scheme by deriving the optimal $ \tilde{M} $ such that the estimation SINR is maximized. 
To start with, we derive the respective power expressions of the useful signal, interference and noise components in the input of the algorithm given in Algorithm \ref{alg: over target parameter estimation}; namely $ \tilde{X}_n[l] $ given in (\ref{eq: tilde Xn[l]}). In doing so, we first recover the accurate signal model without neglecting the intra-symbol Doppler impact. By using the accurate signal model in
(\ref{eq: xn[m] accurate}), in contrast to using (\ref{eq: xn[m] stop and hop}) in Section \ref{sec: sensing framework preprocessing}, the signal in (\ref{eq: tilde xn[m] rewritten}) can be rewritten as 
\begin{align} \label{eq: tilde xn[m] accurate idft}
	& \tilde{x}_n[m] 
	=  
		\sum_{p=0}^{P-1}\overbrace{\tilde{\alpha}_p e^{\mj 2\pi\nu_p n\tilde{M} T_{\mathrm{s}}}}^{\breve{\alpha}_p}  \left( \sum_{l=0}^{\bar{M}-1}\tilde{S}_n[l] e^{-\mj \frac{2\pi l l_p }{\bar{M}}} \myDFT{\bar{M}}{-lm} \right)\times 
\nonumber\\
	& ~~~~~~~~~~~~~~~~ e^{\mj 2\pi\nu_p m T_{\mathrm{s}}}  + z^{\mathrm{A}}_n[m] + z^{\mathrm{B}}_n[m] + w_n[m],
\end{align}
where, to account for the inter-carrier interference (ICI), the time-domain signal $ \tilde{s}_n[\cdot] $  is replaced by its inverse IDFT, as enclosed in the round brackets. Strictly speaking, $ z^{\mathrm{A}}_n[m] $ and $ z^{\mathrm{B}}_n[m]
$ are different from those given in (\ref{eq: zA[n]}) and (\ref{eq: zB[n]}), since $ \tilde{s}_{n-1}\left[ m+\tilde{M}- l_p  \right] $ and $ \tilde{s}_{n}\left[ m+\bar{M} \right] $ are multiplied by extra exponential terms in a point-wise manner. The multiplication, however, changes neither the white Gaussian nature of $ z^{\mathrm{A}}_n[m] $ and $ z^{\mathrm{B}}_n[m] $ nor their variances. This can be validated based on Appendix \ref{app: proof of lemma snm and znm uncorrelated}. Thus, we continue using the same symbols here. 
{Based on the new expression of $ \tilde{x}_n[m]  $, $ \tilde{X}_n[l] $, as given in (\ref{eq: tilde Xn[l]}), 
can be rewritten as in (\ref{eq: tilde Xn[l] ici}), where the coefficient $ \frac{1}{\bar{M}} $ in $ \tilde{X}_n^{\mathrm{S}}[l] $ comes from the product of $ \myDFT{\bar{M}}{-ml}\myDFT{\bar{M}}{ml} $ and the coefficient $ \frac{1}{\sqrt{\bar{M}}} $ in $ \tilde{X}_n^{\mathrm{I}}[l] $ is likewise produced. 
For the same reason that $ z^{\mathrm{A}}_n[m] $ and $ z^{\mathrm{B}}_n[m] $ are reused in (\ref{eq: tilde xn[m] accurate idft}), we continue using $ Z^{\mathrm{A}}_n[l] $, $ Z^{\mathrm{B}}_n[l] $ and $ \tilde{W}_n[l] $ in (\ref{eq: tilde Xn[l] ici}).}
Next, we first derive the power expressions of the four components in (\ref{eq: tilde Xn[l] ici}) and then study their relations. 

\begin{figure*}[!t]
	\begin{align} \label{eq: tilde Xn[l] ici}
		\tilde{X}_n[l] 
		= \frac{\sum_{m=0}^{\bar{M}-1}\tilde{x}_n[m]\myDFT{\bar{M}}{ml}}{\mk \tilde{S}_n[l]}
		= & 
		\underbrace{\sum_{p=0}^{P-1} \frac{\breve{\alpha}_p}{\mk} e^{-\mj \frac{2\pi l l_p }{\bar{M}}} \frac{1}{\bar{M}}\sum_{m=0}^{\bar{M}-1}e^{\mj 2\pi\nu_p m T_{\mathrm{s}}}}_{\tilde{X}_n^{\mathrm{S}}[l] }  
		+\underbrace{\sum_{p=0}^{P-1} \breve{\alpha}_p \sum_{m=0}^{\bar{M}-1}\sum_{\substack{l'=0\\l'\ne l}}^{\bar{M}-1}\frac{\tilde{S}_n[l']}{\mk \tilde{S}_n[l]} e^{-\mj \frac{2\pi l' l_p }{\bar{M}}} e^{\mj 2\pi\nu_p m T_{\mathrm{s}}} \frac{\myDFT{\bar{M}}{(l-l')m}  }{\sqrt{\bar{M}}}}_{\tilde{X}_n^{\mathrm{I}}[l]} \nonumber\\[-5mm]
		&~~~~~~	+ \overbrace{Z^{\mathrm{A}}_n[l]/(\mk \tilde{S}_n[l]) + Z^{\mathrm{B}}_n[l]/(\mk \tilde{S}_n[l])}^{\tilde{X}_n^{\mathrm{Z}}[l]} +  \tilde{W}_n[l] 
	\end{align}
	\vspace{-10mm}
\end{figure*}

Some handy features that will be frequently used later are introduced first.  
We notice that the $ p $-related summation is involved in $ \tilde{X}_n^{\mathrm{S}}[l] $ and $ \tilde{X}_n^{\mathrm{I}}[l] $. Lemma \ref{lm: indepenndet alpha_p} is helpful in simplifying the calculations of their powers. We also notice that the ratio of complex Gaussian variables appears in $ \tilde{X}_n^{\mathrm{I}}[l] $ and $ \tilde{X}_n^{\mathrm{Z}}[l] $. 
The finite moments of such ratio does not exist \cite{book_simon2007probability}. Thus, we provide Proposition \ref{pp: variance of complex normal ratio} to approximate the variance of the ratio. The proof of the proposition is given in Appendix \ref{app: proof of proposition on variance of a complex gaussian ratio}.

\mySpaceTwoMM
\begin{lemma} \label{lm: indepenndet alpha_p}
	\it 
	Let $ \mathcal{X} $ denote a general expression.
	We have 
	\[\myVar{\sum_{p=0}^{P-1}\alpha_p\mathcal{X}}=\sum_{p=0}^{P-1} \sigma_p^2 \myExp{\mathcal{X}^*\mathcal{X}},\]
	where $ \myVar{x} $ denotes the variance of $ x $.
\end{lemma}
\begin{IEEEproof}
	As illustrated in Remark \ref{rmk: basic assumptions}, we consider i.i.d. zero-mean $ \alpha_{p} ~(\forall p) $. Thus, $ \myExp{\sum_{p=0}^{P-1}\alpha_p\mathcal{X}}=0 $. Then, the variance can be calculated as 
	\begin{align}
		& \myVar{\sum_{p=0}^{P-1}\alpha_p\mathcal{X}}= \myExp{ \left(\sum_{p=0}^{P-1}\alpha_p\mathcal{X}\right)\left(\sum_{p=0}^{P-1}\alpha_p\mathcal{X}\right)^* }\nonumber\\
		& = \sum_{p=0}^{P-1} \sigma_p^2 \myExp{\mathcal{X}^*\mathcal{X}},
	\end{align} 
	where the cross-terms are suppressed due to $ \myExp{\alpha_{p_1}\alpha_{p_2}^*}=0 $ given $ \forall p_1\ne p_2 $.
\end{IEEEproof}

\mySpaceTwoMM

	\begin{proposition} \label{pp: variance of complex normal ratio}
		\it Let $ x\sim\mathcal{CN}(0,\sigma_x^2) $ and $ y\sim\mathcal{CN}(0,\sigma_y^2) $ denote two uncorrelated complex Gaussian variables.
		Defining $ z=\frac{x}{y} $, we have $ \myExp{z}=0 $.
		Provided that $ \myProb{|y|\le 1}=\epsilon $, $ \sigma_x^2\ll \sigma_y^2 $ and $ \rho=\sigma_x^2/\sigma_y^2 $, the variance of $ z $ can be approximated by
		\begin{align} \label{eq: signa z^2 result}
			\sigma_{z}^2\approx \mb(\epsilon) \rho,~\mathrm{s.t.}~\mb(\epsilon) = 2\left(  \ln\left( \frac{2(1-\epsilon)}{\sqrt{\epsilon(2-\epsilon)}} \right) - 1  \right).
		\end{align}
	\end{proposition}

\subsection{Power Expressions} \label{subsec: power expressions}

\textit{Power of $ \tilde{X}_n^{\mathrm{S}}[l] $:} 
Applying Lemma \ref{lm: indepenndet alpha_p}, the variance of $ \tilde{X}_n^{\mathrm{S}}[l] $ given in (\ref{eq: tilde Xn[l] ici}), as denoted by $ \sigma_{\mathrm{S}}^2 $, can be calculated as
\begin{align}\label{eq: sigma_S^2}
	& \sigma_{\mathrm{S}}^2 
	=  \frac{1}{\mk^2\bar{M}^2}
	\underbrace{
		\sum_{p=0}^{P-1}\sigma_p^2\myExp{ 
		\begin{array}{l}
			\left( \sum_{m_1=0}^{\bar{M}-1}  e^{\mj 2\pi\nu_{p} m_1 T_{\mathrm{s}}}\right)\\
			~~~~~\times
			\left( \sum_{m_2=0}^{\bar{M}-1}  e^{-\mj 2 \pi \nu_{p} m_2 T_{\mathrm{s}}}\right)
		\end{array}
	}}_{\Pi}  \nonumber\\
	 & \approx  \frac{1 }{ \mk^2 } \Big( \sigma_P^2+\ma - \ma \bar{M}^2 \Big)
	 ,\nonumber\\
	 &\mathrm{s.t.}~\ma=\frac{\pi^2 T_{\mathrm{s}}^2}{3}\sum_{p=0}^{P-1}\sigma_p^2 \nu_p^2,~\sigma_P^2=\sum_{p=0}^{P-1}\sigma_p^2 
\end{align}
where the approximation is due to the Taylor series used for calculating $ \Pi $, as illustrated in Appendix \ref{app: calculating Pi}.

\mySpaceTwoMM

\textit{Power of $ \tilde{X}_n^{\mathrm{I}}[l] $:} Applying Lemma \ref{lm: indepenndet alpha_p}, the variance of $ \tilde{X}_n^{\mathrm{I}}[l] $ given in (\ref{eq: tilde Xn[l] ici}), as denoted by $ \sigma_{\mathrm{I}}^2 $, can be calculated as 
\begin{align} \label{eq: sigma_I^2}
	& \sigma_{\mathrm{I}}^2 
	=  \myExp{\left| \frac{\tilde{S}_n[l']}{\mk \tilde{S}_n[l]} \right|^2}\times 
	\nonumber\\
	& \underbrace{\sum_{p=0}^{P-1} \sigma_p^2 \myExp{ \begin{array}{l}
			\sum_{{l'=0,l'\ne l}}^{\bar{M}-1} \left( 
				\sum_{m_1=0}^{\bar{M}-1}  e^{\mj 2\pi\nu_{p} m_1 T_{\mathrm{s}}}
				\frac{\myDFT{\bar{M}}{(l-l')m_1}  }{\sqrt{\bar{M}}}
			\right)\\
			~~~~~\times
			\left( \sum_{m_2=0}^{\bar{M}-1}  e^{-\mj 2 \pi \nu_{p} m_2 T_{\mathrm{s}}}\frac{\myDFT{\bar{M}}{-(l-l')m_2}  }{\sqrt{\bar{M}}} \right)
	\end{array} }}_{\Gamma}, \nonumber\\
& = 
{ {\mb(\epsilon)\mk^{-2}}(\ma\bar{M}^2 - \ma)},~\mathrm{s.t.}~\ma=\frac{\pi^2 T_{\mathrm{s}}^2}{3}\sum_{p=0}^{P-1}\sigma_p^2 \nu_p^2
\end{align}
where $ \myExp{\left| \frac{\tilde{S}_n[l']}{\mk \tilde{S}_n[l]} \right|^2} = \mb(\epsilon)\mk^{-2} $ according to Lemma \ref{lm: variance of Snl ZAnl Zbnl Wnl} and Proposition \ref{pp: variance of complex normal ratio}, and $ \Gamma $ is calculated in (\ref{eq: Gamma}) (at the top of next page). 
\begin{figure*}[!t]
	\begin{align} \label{eq: Gamma}
	& \Gamma = \sum_{p=0}^{P-1}\sigma_p^2    
			\sum_{{l'=0,l'\ne l}}^{\bar{M}-1}\sum_{m_1=0}^{\bar{M}-1} \sum_{m_2=0}^{\bar{M}-1} e^{\mj 2\pi\nu_{p} (m_1-m_2) T_{\mathrm{s}}}\frac{\myDFT{\bar{M}}{(l-l')(m_1-m_2)}}{\bar{M}^{\frac{3}{2}}}  
\myEqualOverset{(a)} \underbrace{\sum_{p=0}^{P-1}\sigma_p^2{\left(   
 		\begin{array}{l}
 			\sum_{{l'=0}}^{\bar{M}-1}\sum_{m_1=0}^{\bar{M}-1} \sum_{m_2=0}^{\bar{M}-1} \\ [1mm]
 			~~e^{\mj 2\pi\nu_{p} (m_1-m_2) T_{\mathrm{s}}} \frac{\myDFT{\bar{M}}{(l-l')(m_1-m_2)}}{\bar{M}^{\frac{3}{2}}}  
 		\end{array}\right)}}_{\Gamma_1} \nonumber\\[-3mm]
& ~~~~~~~~~~ -
\underbrace{\sum_{p=0}^{P-1}\sigma_p^2\left(   
\begin{array}{l}
	\sum_{{l'=l}}\sum_{m_1=0}^{\bar{M}-1} \sum_{m_2=0}^{\bar{M}-1} \\ [1mm]
	~~e^{\mj 2\pi\nu_{p} (m_1-m_2) T_{\mathrm{s}}} \frac{\myDFT{\bar{M}}{0\times (m_1-m_2)}}{\bar{M}^{\frac{3}{2}}}  
\end{array}\right)}_{{\Pi}/{\bar{M}^2}}  
 \myEqualOverset{(b)} \sigma_P^2 - \Big( \sigma_P^2+\ma-\ma\bar{M}^2 \Big) = \ma\bar{M}^2 - \ma.
\end{align}
\vspace{-5mm}
\end{figure*}
In the calculation, the result $ \myEqualOverset{(a)} $ is obtained by separately considering the summation over $ l'=0,1,\cdots,\bar{M}-1 $ (leading to $ \Gamma_1 $) and the special case of $ l'=l $ (which happens to become $ \Pi/\bar{M}^2 $). As calculated in Appendix \ref{app: deriving Gamma_1=1}, $ \Gamma_1=\sigma_P^2 $. Then, combining 
the result of $ \Pi $ calculated in Appendix \ref{app: calculating Pi}, we obtain the result $ \myEqualOverset{(b)} $ in (\ref{eq: Gamma}).

\mySpaceTwoMM

\textit{Power of $ \tilde{X}_n^{\mathrm{Z}}[l] $:} 
According to Lemma \ref{lm: variance of Snl ZAnl Zbnl Wnl}, $ Z^{\mathrm{A}}_n[l] $, $ Z^{\mathrm{B}}_n[l] $ and 
$ \tilde{S}_n[l] $ are zero-mean complex Gaussian variables. Thus, applying Lemma \ref{pp: variance of complex normal ratio}, we have 
\begin{align} \nonumber
	\myVar{ \frac{Z^{\mathrm{A}}_n[l]}{\mk \tilde{S}_n[l]} }=&\myExp{\left| \frac{Z^{\mathrm{A}}_n[l]}{\mk \tilde{S}_n[l]} \right|^2} = \frac{\mb(\epsilon)\sigma_{Z^{\mathrm{A}}}^2}{\mk^2\sigma_{d}^2} \nonumber\\
	=& \frac{\mb(\epsilon)\mk^{-2}\imax \sum_{p=0}^{P-1}(p+1)\sigma_p^2}{P\bar{M}}, \nonumber
\end{align}
where $ \sigma_{Z^{\mathrm{A}}}^2 $ is given in (\ref{eq: sigma ZA ^2}). Likewise, we obtain
\begin{align} \nonumber
	\myVar{ \frac{Z^{\mathrm{B}}_n[l]}{\mk \tilde{S}_n[l]} }= \frac{\mb(\epsilon)\mk^{-2}(Q-\imin)\sum_{p=0}^{P-1}(P-p)\sigma_p^2}{P\bar{M}}. \nonumber
\end{align}
Using the above two results, the variance of $ \tilde{X}_n^{\mathrm{Z}}[l] $ given in (\ref{eq: tilde Xn[l] ici}), as denoted by $ \sigma_{\mathrm{Z}}^2 $, can be calculated as 
\begin{align} \label{eq: sigma_Z^2}
	& \sigma_{Z}^2  = \myVar{ \frac{Z^{\mathrm{A}}_n[l]}{\mk \tilde{S}_n[l]} } +  \myVar{ \frac{Z^{\mathrm{B}}_n[l]}{\mk \tilde{S}_n[l]} } + \nonumber\\ 
	& ~~~~~~~~ \myExp{\left(\frac{Z^{\mathrm{A}}_n[l]}{\mk \tilde{S}_n[l]}\right)^*\frac{Z^{\mathrm{B}}_n[l]}{\mk \tilde{S}_n[l]}} + \myExp{\left(\frac{Z^{\mathrm{B}}_n[l]}{\mk \tilde{S}_n[l]}\right)^*\frac{Z^{\mathrm{A}}_n[l]}{\mk \tilde{S}_n[l]}} \nonumber\\
	& \approx \frac{\mb(\epsilon)\left(
		\begin{array}{l}
			\imax \sum_{p=0}^{P-1}(p+1)\sigma_p^2		 \\
			~~~~~~~~~+ (Q-\imin)\sum_{p=0}^{P-1}(P-p)\sigma_p^2		
		\end{array}
	\right)}{\mk^2P\bar{M}}\nonumber\\
& \overset{\substack{\imin=0\\\imax=Q}}{\le} ~~\frac{\mb(\epsilon)Q(P+1)\sigma_P^2}{\mk^2P\bar{M}},~\mathrm{s.t.}~\sigma_P^2 = \sum_{p=0}^{P-1}\sigma_p^2,
\end{align}
where the two expectations are approximately zeros, as proved in Appendix \ref{app: E{ZA/S ZB/S}=0}. The equality in the last line can be achieved when the two conditions are satisfied, $  \imin=0$ and $ \imax=Q $. 
These conditions are likely to hold for a large $ P $, since the larger $ P $ is, the more variety can present among targets.

\mySpaceTwoMM

\textit{Power of $ \tilde{W}_n[l]  $:} According to (\ref{eq: tilde Xn[l]}), we have $ \tilde{W}_n[l]=W_n[l]/(\mk \tilde{S}_n[l])  $ which is the ratio of two complex Gaussian. Thus, directly applying Proposition \ref{pp: variance of complex normal ratio} gives us 
\begin{align} \label{eq: sigma_W^2}
	\sigma_W^2 = \myVar{\tilde{W}_n[l]} = \frac{\mb(\epsilon)\sigma_w^2}{\mk^2\sigma_d^2},
\end{align}
where $ \myVar{W_n[l]} = \sigma_w^2 $, as proved in Lemma \ref{lm: variance of Snl ZAnl Zbnl Wnl}.

\subsection{Maximizing SINR for Parameter Estimation}
With the power of individual signal components in $ {\tilde{X}_n[l]} $ derived, we proceed to analyze the overall SINR for the proposed parameter estimation method and derive the optimal $ \tilde{M} $. 
Note that 
a two-dimensional DFT along $ n $ and $ l $ is taken for $ {\tilde{X}_n[l]} $ before estimating target parameters; see Algorithm \ref{alg: over target parameter estimation}. 
The maximum SINR improvement brought by the DFT is $ \bar{M}\tilde{N} $; see (\ref{eq: breve X_{tilde n}^{(p)} (tilde l)}). This is because $ \tilde{X}_n^{\mathrm{S}}[l] $ is coherently accumulated with the maximum amplitude gain of $ \bar{M}\tilde{N} $, while $ \tilde{X}_n^{\mathrm{I}}[l] $ and $ \tilde{X}_n^{\mathrm{Z}}[l] $ are only incoherently accumulated with the power gain of $ \bar{M}\tilde{N} $. 
With the SINR gain of $ \bar{M}\tilde{N} $
taken into account, the overall SINR for parameter estimation can be given by
\begin{align} \label{eq: gamma overall SINR}
	\gamma = \frac{f(\bar{M})}{g(\bar{M})} \myDef \frac{\bar{M}\tilde{N}\sigma_{\mathrm{S}}^2}{\sigma_{\mathrm{I}}^2 + \sigma_{{Z}}^2 + \sigma_{{W}}^2},
\end{align}
where $ \sigma_{\mathrm{S}}^2 $, $ \sigma_{\mathrm{I}}^2 $, $ \sigma_Z^2 $ and $ \sigma_W^2 $ are derived in (\ref{eq: sigma_S^2}), (\ref{eq: sigma_I^2}), (\ref{eq: sigma_Z^2}) and (\ref{eq: sigma_W^2}), respectively.
To avoid overly cumbersome expressions, we denote the numerator and denominator as different function of $ \bar{M} $ and study their monotonic features separately. 
In particular, the following interesting results are achieved, with the proof given in Appendices \ref{app: proof of proposition on concave numerator function} and \ref{app: proof of proposition on convex denominator in sinr}.

\mySpaceTwoMM

\begin{proposition} \label{pp: concave numerator in sinr}
	\it 
	The numerator function $ f(\bar{M}) $ in (\ref{eq: gamma overall SINR}) is concave w.r.t. $ \bar{M} $. Given $ \bar{M}\gg 3Q/2 $, $ f(\bar{M}) $ is maximized at 
	\[\bar{M}=\bar{M}_f^* \approx \sqrt[3]{{(Q\ma + Q\sigma_P^2)}\big/{(2\ma)}}.\]
\end{proposition}

\begin{proposition}\label{pp: convex denominator in sinr}
	\it 
	The denominator function $ g(\bar{M}) $ in (\ref{eq: gamma overall SINR}) is convex w.r.t. $ \bar{M} $ and is minimized at
	\[\bar{M}=\bar{M}_g^* = \sqrt[3]{{(Q(P+1)\sigma_P^2)}\big/{(2\ma P)}}.\]
\end{proposition}

\mySpaceTwoMM

From the two propositions, $ f(\bar{M}) $
and $ g(\bar{M}) $ seem to achieve the extremum at different values of $ \bar{M} $. Nevertheless, taking into account the practical values of the variables in their respective expressions, we have the following result.

\mySpaceTwoMM

\begin{corollary}\label{col: Mf^*approx Mg^*}
	\it The optimal $ \bar{M} $ for the numerator and denominator functions in (\ref{eq: gamma overall SINR}) satisfy $ \bar{M}_f^* \approx \bar{M}_g^* \approxeq\sqrt[3]{{ Q\sigma_P^2}\big/{(2\ma)}} $, where the equality can be approached as $ P $ becomes larger.
\end{corollary}

\mySpaceTwoMM

\begin{IEEEproof}
	As defined in (\ref{eq: sigma_S^2}), $ \ma=\frac{\pi^2 T_{\mathrm{s}}^2}{3}\sum_{p=0}^{P-1}\sigma_p^2 \nu_p^2 $. Thus we have the following upper bound of $ 2\ma $	
	\[2\ma =\frac{\pi^2 T_{\mathrm{s}}^2}{3}\sum_{p=0}^{P-1}2\sigma_p^2 \nu_p^2\le \frac{\pi^2 T_{\mathrm{s}}^2}{3}\sum_{p=0}^{P-1} (\sigma_p^2 +  \nu_p^2).\]
	Applying the above inequality, we have 
	\[
	Q\sigma_P^2/(2\ma)  \ge \frac{Q}{\frac{\pi^2 T_{\mathrm{s}}^2}{3} (1 +  \frac{1}{\sigma_P^2}\sum_{p=0}^{P-1}\nu_p^2)}\gg Q/2
	\]
	where the last result is because $ T_{\mathrm{s}} $ can be in the order of $ 10^{-6} $ while $ \nu_p $ is only in the order of $ 10^3 $. Reflecting the above relation in Proposition \ref{pp: concave numerator in sinr}, we obtain $ \bar{M}_f^* \approx \sqrt[3]{{ Q\sigma_P^2}\big/{(2\ma)}} $. 	
	Given a large $ P $, $ \bar{M}_g^* $ derive in Proposition \ref{pp: convex denominator in sinr} becomes $ \bar{M}_g^*\approxeq \sqrt[3]{{ Q\sigma_P^2}\big/{(2\ma)}}\approx\bar{M}_f^* $.
\end{IEEEproof}

\mySpaceTwoMM

In summary, we highlight below the key features of the proposed OTFS sensing methods. \textit{First}, the actual
range and Doppler dimensions, as denoted by $ \tilde{M} $ and $ \tilde{N} $, respectively, are not limited to the original dimensions of the underlying OTFS system, i.e., $ M $ and $ N $.
\textit{Second}, the maximum range that can be estimated by the proposed design is not subject to the CP length of the OTFS system and can be flexibly adjusted by changing $ Q $, i.e., the length of the proposed VCP. \textit{Third}, the SINR for target parameter estimation can be maximized by setting
\begin{align} \label{eq: tilde M optimal}
	\tilde{M}=\sqrt[3]{\frac{ Q\sigma_P^2}{(2\ma)}} + Q,~\tilde{N} = \myRound{\frac{MN}{\tilde{M}}}
\end{align}
where $ \ma $ and $ \sigma_P^2 $ are defined in (\ref{eq: sigma_S^2}). Note that the relation $ \tilde{M}=\bar{M}+Q $, as given in (\ref{eq: tilde sn[m]}), is used above. \textbb{Also note that the optimization in this section has included the potential ICI caused by the intra-sub-block Doppler impact. Thus, even if a large $ Q $ results in a large $ \tilde{M} $ according to (\ref{eq: tilde M optimal}), the optimal $ \tilde{M} $ is guaranteed to maximizes the sensing SINR, whether ICI is present or not.}

\section{Simulation Results} \label{sec: simulations}

In this section, simulation results are provided to validate the 
proposed designs and analyses. 

\subsection{Comparison with Benchmark Method}

In a first set of simulations, we compare the proposed design with the state-of-the-art OTFS sensing method \cite{OTFS_jcas2020twc} which performs the maximum likelihood (ML) estimation. In short, the ML method \cite{OTFS_jcas2020twc} first transforms $ x[i] $ given in (\ref{eq: x[i]}) into the delay-Doppler domain by performing an inverse symplectic Fourier transform, i.e., 
the inverse transform of that performed in (\ref{eq: S[m,n]}). The result is stacked into an $ I $-dimensional $ (I=MN) $ column vector, as denoted by $ \mathbf{y}_{\mathrm{ML}} $. Then a multi-layer two-dimensional searching is carried out, using the coefficients $ \mathbf{H}_{\mathrm{ML}}^a(l_g^a,\nu_g^a) \mathbf{d}  $, where $ \mathbf{d} $ is obtained by stacking $ d_{kM+l} $ into an $ I $-dimensional column vector ($ k $ first) and $ \mathbf{H}_{\mathrm{ML}}^a(l_g^a,\nu_g^a) $ is the input-output response matrix at a delay-Doppler grid $ (l_g^a,\nu_g^a) $; refer to \cite[Eq. (15)]{OTFS_jcas2020twc} for its detailed expression. The superscript $ ()^a $
denotes the layer index and the subscript $ ()_g $ denotes the grid index. 
For a single target, the 2D searching can be depicted as follows,
\begin{align}\label{eq: ML benchmark}
	(\hat{l}^a,\hat{\nu}^a):~\max_{(l_g^a,\nu_g^a)\in \mathcal{S}_a} \frac{\left(\mathbf{H}_{\mathrm{ML}}^a(l_g^a,\nu_g^a) \mathbf{d}\right)^{\mathrm{H}} \mathbf{y}_{\mathrm{ML}}}{\left( \left(\mathbf{H}_{\mathrm{ML}}^a(l_g^a,\nu_g^a) \mathbf{d}\right)^{\mathrm{H}}
		\left(\mathbf{H}_{\mathrm{ML}}^a(l_g^a,\nu_g^a) \mathbf{d}\right)
		\right)}, 
\end{align}
where $ (\hat{l}^a,\hat{\nu}^a) $ is the pair of estimated parameters and $ \mathcal{S}_{a} $ is the delay-Doppler region to be searched at layer $ a $. 

For simulation efficiency, we take $ {l}_g^a=l^a~(\forall a,g) $, where $ l^a $ is the true value of sample delay at layer $ a $. Thus, we only perform ML estimation for the target velocity. Moreover, we set the number of grids in each layer as five, i.e., $ |\mathcal{S}_a|=5 $, and the total number of layers as eight, i.e., $ a=0,1,\cdots,7 $. More specifically, at $ a=0 $, we set $ \nu_g^0=g\df/4~(g=0,1,2,3,4) $. Note that $ \df $, the sub-carrier interval, is also the unambiguously measurable region of the Doppler frequency for the ML method \cite{OTFS_jcas2020twc}.
For $ a\ge 1 $, the searching region of layer $ a $ is reduced to $ \df/4^a $ centered at $ \hat{\nu}^{a-1} $. In addition, we set a rule that stops the iterative searching at layer $ a $ if $ \hat{\nu}_g^a\ne {\nu}^a $ and take $ {\nu}^a $ as the final velocity estimate, where $ {\nu}^a $ is the true value. 
For fair comparison with the ML method \cite{OTFS_jcas2020twc}, the key system parameters set therein are used here only with minor modification. The settings are summarized in Table \ref{tab: simulation parameters benchmark}.

\begin{table}[!t]\footnotesize
	\captionof{table}{Simulation Parameters}
	\vspace{-3mm}
	\begin{center}
		\begin{tabular*}{88mm}{c| l|l }
			\hline
			Variable &
			Description &
			Value 
			\\
			\hline
			$ f_{\mathrm{c}} $    
			&  Carrier frequency 
			&	$ 5.89 $ GHz 
			\\				
			\hline
			$ B $ 
			& Bandwidth 
			& $ 10 $ MHz
			\\				
			\hline
			$ \df $    
			&  Sub-carrier interval
			&	$ B/M $ MHz 
			\\				
			\hline
			$ M $ 
			& No. of sub-carriers per symbol
			& $ 25 $
			\\				
			\hline
			$ N $ 
			& No. of symbols
			& $ 40 $
			\\				
			\hline	
			$ \sigma_d^2 $ 
			& Power of data symbol $ d_{kM+l} $ give in (\ref{eq: S[m,n]})
			& $ 0 $ dB
			\\				
			\hline
			$ \sigma_0^2 $ 
			& Power of $ \alpha_p~(p=0) $; see (\ref{eq: xn[m]})
			& $ 0 $ dB
			\\				
			\hline
			$ \sigma_w^2 $ 
			& Variance of AWGN $ w_n[m] $ given in (\ref{eq: xn[m]})
			& $ -10:2:10 $ dB
			\\				
			\hline
			$ r $ 
			& Target range
			& $ 30 $ m
			\\				
			\hline
			$ v $ 
			& Target velocity
			& $ 80 $ km/h
			\\				
			\hline
			$ N_{\mathrm{iter}} $ 
			& No. of iteration for Algorithm \ref{alg: estimating delta_p}
			& 5
			\\				
			\hline
		\end{tabular*}
		\vspace{-3mm}
	\end{center}
	\label{tab: simulation parameters benchmark}
\end{table}

\begin{figure}[!t]
	\centering 	
	\includegraphics[width=80mm]{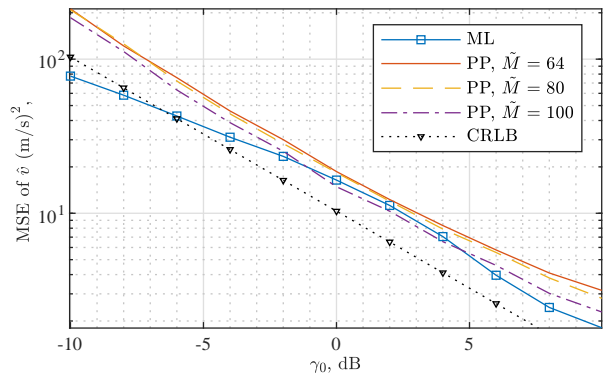}
	\caption{MSE of velocity estimation, denoted by $ \hat{v} $, versus $ \gamma_0(=\sigma_0^2/\sigma_w^2) $, i.e., the SNR of $ x_n[m] $ given in (\ref{eq: xn[m]}), where $ ML $ is for the benchmark method \cite{OTFS_jcas2020twc} and PP for the proposed design.}
	\label{fig: mse vs snr 80kmh diff tilde M}
\end{figure}

Fig. \ref{fig: mse vs snr 80kmh diff tilde M} plots the mean squared error (MSE) of the velocity estimation, denoted by $\hat{v} $, against the SNR, denoted by $ \gamma_0 $. Here, $ \gamma_0=(\sigma_0^2/\sigma_w^2) $ is defined based on the sensing-received signal in the time domain, i.e., $ x_n[m] $ given in (\ref{eq: xn[m]}). 
We see that, in overall, the proposed method has a comparable performance to the ML method \cite{OTFS_jcas2020twc}. 
Interestingly, we see different performance relation between the two methods in three SNR regions. 

\begin{enumerate} [leftmargin=*]
	\item As $ \gamma_0 $ increases from $ -10 $ dB to $ -2 $ dB, 
	the ML outperforms the proposed method, with the performance gap becoming increasingly smaller. This is because the proposed estimation algorithm, as summarized in Algorithm \ref{alg: estimating delta_p}, can return any value in $ [0,\frac{1}{\tilde{M}T_{\mathrm{s}}}] $\footnote{In Section \ref{sec: sensing framework parameter estimation}, the estimation of Doppler frequency is turned into estimating $ n_p(=\tilde{N}\tilde{M}T_{\mathrm{s}}\nu_p) $ defined in (\ref{eq: breve X_{tilde n}^{(p)} (tilde l)}). The proposed algorithm for estimating $ n_p $, as summarized in Algorithm \ref{alg: estimating delta_p}, can return any value in $ [0,\tilde{N}] $ as the estimate of $ n_p $ in the presence of strong noises. Thus, the estimation of $ nu_p $ can take any value in the region of $[0,\frac{1}{\tilde{M}T_{\mathrm{s}}}]  $.}, in the presence of strong noises. In contrast, the way the ML method is simulated makes its velocity estimation error bounded to $ 80/3.6 $;
	
	\item \textbb{For $ \gamma_0\in [-1,5] $ dB, the proposed method is able to outperform the ML method. The reason is that the proposed method provides off-grid estimations, while the ML method performs on-grid searching}; and
	
	\item As $ \gamma_0 $ increases over $ 5 $ dB, an increasing (yet small) performance gap between the proposed and the ML methods can be seen. The slight degrading of the proposed method is caused by the interference introduced during adding VCP; refer to Fig. \ref{fig: add VCP}. In particular, when the noise power keep decreasing, the impact of the inter- and intra-symbol interference becomes gradually dominant. 
\end{enumerate}
Nevertheless, we see from Fig. \ref{fig: mse vs snr 80kmh diff tilde M} that increasing $ \tilde{M} $ can help reduce the impact of the interference. 
This is expected. 
Based on the parameter setting in Table \ref{tab: simulation parameters benchmark} and Corollary \ref{col: Mf^*approx Mg^*}, we have $ \bar{M}_f^* \approx \bar{M}_g^* \approxeq 247.8675 $. According to Propositions \ref{pp: concave numerator in sinr} and \ref{pp: convex denominator in sinr}, the estimation SINR increases with $ \tilde{M} $ when $  \tilde{M}\le \bar{M}_f^* $. That is, a larger $ \tilde{M} $ corresponds to a more accurate $ \hat{v} $ estimation.

\textbb{Fig. \ref{fig: mse vs snr 80kmh diff tilde M} also plots the CLRB of velocity estimation to validate the high accuracy of the parameter estimation methods developed in Section \ref{sec: sensing framework parameter estimation}. 
As derived Appendix \ref{app: crlb}, the CRLB is given by 
\begin{align} \label{eq: crlb velocity estimation}
	\mathrm{CRLB}\{\hat{v}\} = \frac{\lambda^2}{4}\frac{1}{(\tilde{M}\tilde{N}T_{\mathrm{s}})^2} \frac{6}{4\pi^4 \tilde{N} \bar{M}\gamma_0/\mb(\epsilon) },
\end{align} 
where $ \mb(\epsilon) $ is given in Proposition \ref{pp: variance of complex normal ratio}. 
We see from Fig. \ref{fig: mse vs snr 80kmh diff tilde M} that the asymptotic performance of our estimation is slightly worse than the ML estimation. This owes to the interference introduced when adding VCP. Note that the low-SNR performance of the ML method is not accurate in Fig. \ref{fig: mse vs snr 80kmh diff tilde M}. As explained above, this is because the way we simulate the method bounds estimation error.}

\begin{figure}[!t]
	\centering 	
	\includegraphics[width=80mm]{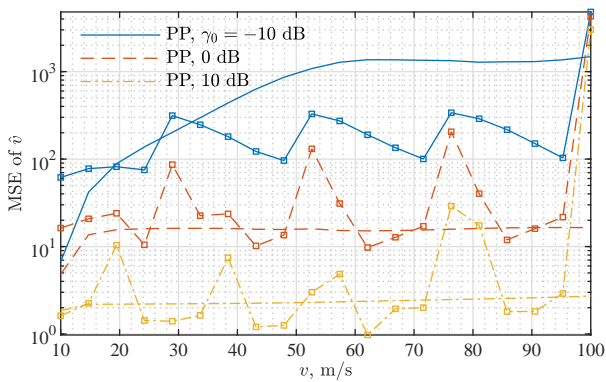}
	\caption{MSE of $ \hat{v} $ versus $ v $, where $ \tilde{M}=100 $ is taken for the proposed method (PP) and the curves with markers correspond to the ML method \cite{OTFS_jcas2020twc}.}
	\label{fig: mse vs velocity diff snr}
\end{figure}

Fig. \ref{fig: mse vs velocity diff snr} plots the MSE of $ \hat{v} $ versus different values of $ v $. We see that the proposed method can achieve a comparable performance to the ML method \cite{OTFS_jcas2020twc} at a moderately high or high SNRs. We also see that the proposed method presents an even performance over a large region of $ v $, while the good performance of the ML method can be velocity-selective. 
From Figs. \ref{fig: mse vs snr 80kmh diff tilde M} and \ref{fig: mse vs velocity diff snr}, we conclude that the proposed method can provide an ML-like estimation performance for OTFS-based sensing.

\subsection{Wide Applicability of Proposed Methods}
Next, we validate the analysis of the impact of $ \tilde{M} $ on the proposed OTFS sensing framework. To also show the wide applicability of the proposed method, we employ another set of system parameters, as summarized in Table \ref{tab: simulation parameters wide applicability}, where $ \mathcal{U}_{[a,b]} $ denotes the uniform distribution in $ [a,b] $. Note that $ P=4 $ means four targets are set. Also note that the Doppler frequency of $ 4.63 $ KHz corresponds to the maximum target velocity of $ 500 $ km/h.

\begin{table}[!b]\footnotesize
	\captionof{table}{Simulation Parameters}
	\vspace{-3mm}
	\begin{center}
		\begin{tabular*}{88mm}{c| l|c | l | c| l}
			\hline
			Variable &
			Value &
			Variable &
			Value &
			Variable &
			Value 
			\\
			\hline
			$ f_{\mathrm{c}} $    
			& $ 5 $ GHz
			&  $ B $ 
			&	$ 12 $ MHz 
			& $ P $
			& $ 4 $
			\\				
			\hline
			$ M $ 
			& $ 400 $ 
			& $ N $
			& $ 100 $
			& $ l_p~(\forall p) $
			& $ \mathcal{U}_{[0,Q-1]} $
			\\				
			\hline
			$ \sigma_w^2 $    
			&  $ -10 $ dB
			&	$ \sigma_d^2 $ 
			& $ 0 $ dB
			& $ \sigma_p~(\forall p) $
			& $ 0 $ dB
			\\				
			\hline
			$ \sigma_0^2 $ 
			& $ 0 $ dB
			& $ \nu_p~(\forall p) $
			& \multicolumn{2}{l}{$ \mathcal{U}_{[-4.63~\mathrm{KHz},4.63~\mathrm{KHz}]} $}
			\\				
			\hline			
		\end{tabular*}
		\vspace{-3mm}
	\end{center}
	\label{tab: simulation parameters wide applicability}
\end{table}	

\begin{figure}[!t]
	\centering 
	
	\begin{minipage}{42mm}
		\includegraphics{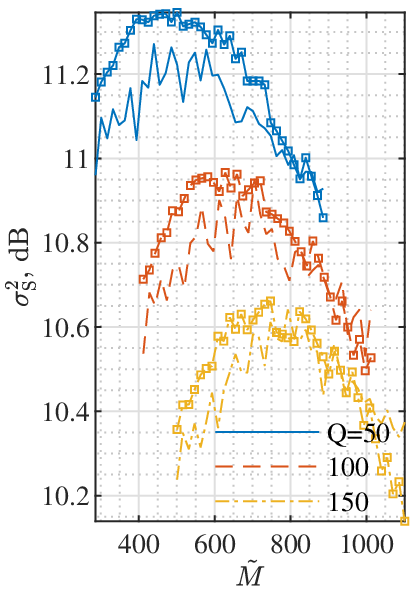}
	\end{minipage}\hfill
	\begin{minipage}{42mm}
		\includegraphics{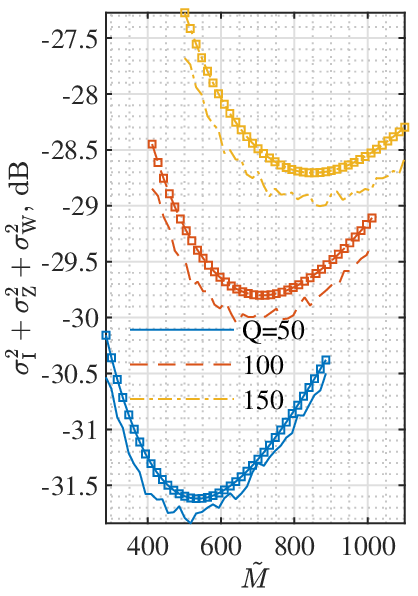}
	\end{minipage}
	\caption{Illustrating the impact of $ \tilde{M} $ on the powers of signal and other components, as derived in Section \ref{subsec: power expressions}, where the curves with markers are theoretical results while the ones without markers are simulated.}
	\label{fig: sinr vs tilde M}
\end{figure}

Fig. \ref{fig: sinr vs tilde M} illustrates the impact of $ \tilde{M} $ on the powers of the signal and other components in $ \tilde{X}_n[l] $ given in (\ref{eq: tilde Xn[l] ici}), where the power expressions of $ \sigma_{\mathrm{S}}^2 $, $ \sigma_{\mathrm{I}}^2 $, $ \sigma_Z^2 $ and $ \sigma_W^2 $, as derived in (\ref{eq: sigma_S^2}), (\ref{eq: sigma_I^2}), (\ref{eq: sigma_Z^2}) and (\ref{eq: sigma_W^2}), respectively, are plotted as the theoretical references. From Fig. \ref{fig: sinr vs tilde M}, we see that the derived expressions can precisely depict the actual powers of different signal components. We also see from the left sub-figure that the signal power presents an approximate concave relation with $ \tilde{M} $, as proved in Proposition \ref{pp: concave numerator in sinr}. Moreover, we see from the right sub-figure that the overall power of interference plus noises is indeed a convex function of $ \tilde{M} $, which validates Proposition \ref{pp: convex denominator in sinr}. 
In addition, jointly comparing the two sub-figures, we can see that the extrema of $ \sigma_{\mathrm{S}}^2 $
 and $ (\sigma_{\mathrm{I}}^2+\sigma_{\mathrm{Z}}^2+\sigma_{\mathrm{W}}^2) $ are achieved at approximately identical $ \tilde{M} $ for each value of $ Q $. This confirms the analysis in Corollary \ref{col: Mf^*approx Mg^*}.

\begin{figure}[!t]
	\centering 
	
	\begin{minipage}{42mm}
		\includegraphics{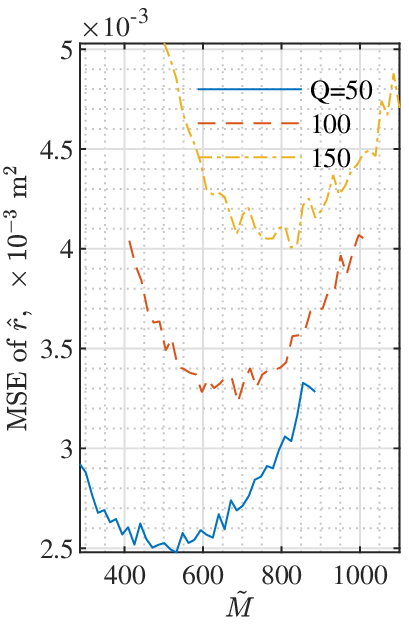}
	\end{minipage}\hfill
	\begin{minipage}{42mm}
		\includegraphics{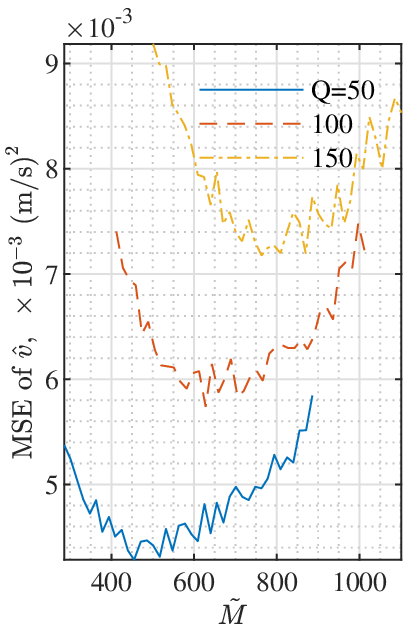}
	\end{minipage}
	\caption{MSE of parameter estimations versus $ \tilde{M} $, where the mean is calculated over the squared estimation errors of all the four targets.}
	\label{fig: mse vs tilde M}
\end{figure}

Fig. \ref{fig: mse vs tilde M} plots the MSE of range and velocity estimations against $ \tilde{M} $. We see that, as consistent with the SINR variations reflected in Fig. \ref{fig: sinr vs tilde M}, the MSEs of the estimations of both parameters are convex functions of $ \tilde{M} $ and are minimized at the optimal $ \tilde{M} $ derived in Corollary \ref{col: Mf^*approx Mg^*}. We also see that the proposed method achieves the high-accuracy estimations of the ranges and velocities of all four targets (as the MSE is calculated over the squared estimation errors of all targets).  
Moreover, we see from Fig. \ref{fig: sinr vs tilde M} that $ Q $ also has a non-trivial impact on OTFS sensing performance. In particular, as $ Q $ becomes larger, the whole MSE curve shifts upwards. 
This is because a larger $ Q $ leads to a smaller number of samples in the essential part of each sub-block; see Fig. \ref{fig: add VCP}; as a further result, 
the coherent accumulation gain of the two-dimensional DFT performed in (\ref{eq: breve X_{tilde n}^{(p)} (tilde l)}) for parameter estimation is also reduced.

\begin{figure}[!t]
	\centering 
	
	\begin{minipage}{42mm}
		\includegraphics{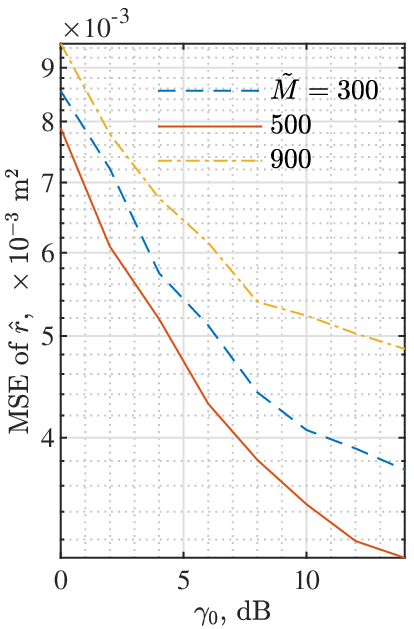}
	\end{minipage}\hfill
	\begin{minipage}{42mm}
		\includegraphics{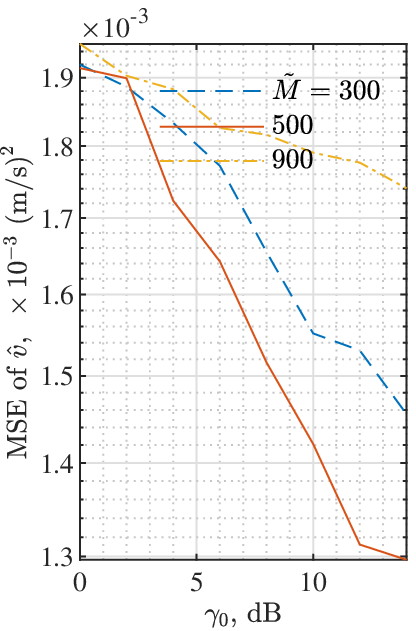}
	\end{minipage}
	\caption{MSE of parameter estimations versus $ \gamma_0 $, where $ Q=50 $ and the mean is calculated over the squared estimation errors of all the four targets.}
	\label{fig: mse vs sinr under diff tilde M}
\end{figure}

Fig. \ref{fig: mse vs sinr under diff tilde M} illustrates the MSE of parameter estimations against $ \gamma_0 $. Based on Fig. \ref{fig: sinr vs tilde M}, we see that the optimal $ \tilde{M} $ is about $ 500 $. Thus, we take three values of $ \tilde{M} $
in this simulation to further illustrate the impact of $ \tilde{M} $ under different $ \gamma_0 $. As consistent with Fig. \ref{fig: sinr vs tilde M} and \ref{fig: mse vs tilde M}, the best MSE performance versus $ \gamma_0 $ is achieved at $ \tilde{M}=500 $. Moreover, we also see that the MSE performance under $ \tilde{M}=500 $ is better than that under $ \tilde{M}=900 $. This is also consistent with what observed in Fig. \ref{fig: mse vs tilde M}. In addition, we see that the impact of $ \tilde{M} $
is more prominent at a higher SNR. This is because as noise becomes weaker, the interference component has a more dominating effect, 
which also makes the impact of $ \tilde{M} $ more obvious.

\section{Conclusions and Remarks}\label{sec:conclusions}

To advocate OTFS-based JCAS for future IIoT as an ultra-reliable, low-latency and high-accuracy communication and sensing solution, we develop in this paper a low-complexity OTFS sensing method with the near-ML performance. This is achieved by a series of waveform pre-processing which addresses the challenges of ICI and ISI, and successfully removes the the impact of communication data symbols in the time-frequency domain without amplifying noises. This is also accomplished by a high-accuracy and off-grid method for estimating range of velocity. The complexity of the method is only dominated by a two-dimensional DFT. This is further attained by the comprehensive analysis of SINR for parameter estimation and the optimization of a key parameter in the proposed method. Extensive simulations are provided, validating the near-ML performance and wide applicability of our method as well as the precision of our SINR analysis.

We remark that additional research effort would be necessary to make OTFS-based JCAS more ready for IIoT applications. 
\textit{First}, sensing between the distributed, non-coherent, asynchronous transmitter and receiver is yet to be solved, where the estimation ambiguities caused by timing and frequency offset are challenging to remove \cite{Kai_rahman2020enablingSurvey}. 
\textit{Second}, networked sensing can combine the capabilities and diversities of different IIoT nodes for better performance, which is yet largely unexplored. \textit{Third}, using sensing results to assist communications can further enhance the energy-efficiency of JCAS. A recent work \cite{OTFS_yuan2021integratedSensingCOmmunicationOTFS} shows the benefit of doing so in the vehicular network. More studies are necessary to account for the unique needs/features of IIoT applications. \textbb{Fourth, in the spirit of JCAS, we may be able to make changes to the transmitted OTFS signal, e.g., inserting more CPs as in OFDM systems. This, however, requires a holistic evaluation on the trade-off between the performance loss of communications and performance gain of sensing.
The state-of-the-art OFDM-based JCAS waveform design \cite{liu2019robust,liu2017multiobjective,liu2017adaptive,liu2020super,liu2019joint} can provide good references for future OTFS JCAS designs. 
}

\appendix
\subsection{Proof of Lemma \ref{lm: sn[m] zn[m] uncorrelated}} \label{app: proof of lemma snm and znm uncorrelated}
\begin{figure*}[!t]
	\begin{align} \label{eq: sigma zA ^2}
		& \sigma_{z^{\mathrm{A}}}^2[m] =\myExp{\left| z^{\mathrm{A}}_n[m] \right|^2} = \myExp{ \left(\sum_{p_1=0}^{P-1}\tilde{\alpha}_{p_1}^* \tilde{s}_{n-1}^*\left[ m+\tilde{M}-i_{p_1} \right] g_{i_{p_1}}[{m}] e^{-\mj 2\pi\nu_{p_1}(n-1)\tilde{M}T_{\mathrm{s}}}\right)  \times  \left(\sum_{p_2=0}^{P-1}\tilde{\alpha}_{p_2} \tilde{s}_{n-1}\left[ m+\tilde{M}-i_{p_2} \right] \right.\right.\nonumber\\
			&\left.\left. g_{i_{p_2}}[{m}] e^{\mj 2\pi\nu_{p_2}(n-1)\tilde{M}T_{\mathrm{s}}}\right) } \overset{(a)}{=} \myExp{ \left(\sum_{p=0}^{P-1}\left|\tilde{\alpha}_{p}\right|^2 \left|\tilde{s}_{n-1}\left[ m+\tilde{M}-i_{p} \right]\right|^2 g_{i_{p}}[{m}] \right) } = \sum_{p=0}^{P-1}\sigma_p^2 \sigma_d^2 g_{i_{p}}[{m}]
	\end{align}
\end{figure*}

As illustrated in Remark \ref{rmk: gaussian of S[m,n]}, $ S[m,n]\sim \mathcal{CN}(0,\sigma_{d}^2) $ is satisfied and independent over $ \forall m,n $. The signal transform in (\ref{eq: s(t)}) resembles an unitary discrete time Fourier transform. Therefore, the resulted signal preserves the white Gaussian feature. Since $ \tilde{s}_n[m] $ is the sampled version of the signal obtained in (\ref{eq: s(t)}), we have $ \tilde{s}_n[m]\sim\mathcal{CN}(0,\sigma_d^2) $ and is independent over $ \forall m,n $. 
Reflecting $ \tilde{s}_n[m]\sim\mathcal{CN}(0,\sigma_d^2)~(\forall n,m) $ in (\ref{eq: zA[n]}) and (\ref{eq: zB[n]}), we see that $ z^{\mathrm{A}}_n[m] $ and $ z^{\mathrm{B}}_n[m] $ are also centered complex Gaussian variables. 

Based on (\ref{eq: zA[n]}), the variance of $ z^{\mathrm{A}}_n[m] $ can be calculated as in (\ref{eq: sigma zA ^2}), where $ \overset{(a)}{=} $ is obtained by suppressing the cross-terms at $ p_1\ne p_2 $. The suppression is enabled by the uncorrelated target scattering coefficients, as illustrated in Remark \ref{rmk: basic assumptions}. We see from (\ref{eq: sigma zA ^2}) that the variance of $ z^{\mathrm{A}}_n[m] $ can change with $ m $ due to the various support of the rectangular function associated with each path, i.e., $ g_{ l_p }[m] $. As shown in Fig. \ref{fig: add VCP}, the number of paths involved in $ z^{\mathrm{A}}_n[m] $ decreases, as $ m $ becomes larger. This leads to the discrete values of $ \sigma_{z^{\mathrm{A}}}^2[m] $, as summarized in Table \ref{tab: value of variance of zAm}. Similar to the above analysis for $ z^{\mathrm{A}}_n[m] $, the variance of $ z^{\mathrm{B}}_n[m] $ can be obtained, as summarized in Table \ref{tab: value of variance of zBm}. 
On the other hand, we notice from Fig. \ref{fig: add VCP} that $ z^{\mathrm{A}}_n[m] $, $ z^{\mathrm{B}}_n[m] $ and $ \tilde{s}_n[m] $ have non-overlapping support. Therefore, the three components are mutually uncorrelated. 

The whiteness of $ z_n^{\mathrm{A}}[0],z_n^{\mathrm{A}}[1],\cdots,z_n^{\mathrm{A}}[\bar{M}-1] $ can be validated based on (\ref{eq: zA[n]}). In particular, we can show that
\begin{align} \label{eq: E{znAm1 znAm2}}
	&\myExp{z_{n}^{\mathrm{A}}[m_1](z_{n}^{\mathrm{A}}[m_2])^*}=\sum_{p=0}^{P-1}\sigma_p^2\times \nonumber\\
	& ~~~~~~~~~~\myExp{\tilde{s}_{n-1}\left[ m_1+\tilde{M}- l_p  \right]  \tilde{s}_{n-1}^*\left[ m_2+\tilde{M}- l_p  \right]},\nonumber
	\\
	& =0,~\forall m_1\ne m_2,
\end{align}
where the rectangular functions in $ z_{n}^{\mathrm{A}}[m_1]$ and $ z_{n}^{\mathrm{A}}[m_2]  $ are dropped for brevity, and the cross-terms are suppressed directly due to $ \myExp{\alpha_{p_1}\alpha_{p_2}^*}=0 $. Similarly, we can validate the whiteness of $ z_n^{\mathrm{A}}[m] $ along the $ n $-dimension given $ \forall m $, and moreover the whiteness of $ z_n^{\mathrm{B}}[m] $ along $ n $- and $ m $-dimensions.

\subsection{Proof of Lemma \ref{lm: variance of Snl ZAnl Zbnl Wnl}} \label{app: proof of lemma on variances of Snl Znl Wnl}

Each component in (\ref{eq: Snl Wnl Znl Gaussian}) is a unitary DFT of the corresponding component given in (\ref{eq: snm zAnm zBnm wnm Gaussian}). Thus, with reference to the proof of Lemma \ref{lm: sn[m] zn[m] uncorrelated} established in Appendix \ref{app: proof of lemma snm and znm uncorrelated}, we can readily show the white Gaussian nature of the four components in (\ref{eq: Snl Wnl Znl Gaussian}). The details are suppressed for brevity. 
Next, we show the derivation of the variance of $ \sigma_{Z^{\mathrm{A}}}^2 $.

Similar to (\ref{eq: tilde Sn[l]}), $ Z^{\mathrm{A}}_n[l] $ can be written as $ 	Z^{\mathrm{A}}_n[l] = \sum_{m=0}^{\imax-1}z_n^{\mathrm{A}}[m] \myDFT{\bar{M}}{ml} $,
where $ \myDFT{\bar{M}}{ml} $ is the DFT basis, as given in (\ref{eq: tilde Sn[l]}). The variance of $ Z^{\mathrm{A}}_n[l] $ can be calculated as 
\begin{align} \label{eq: sigma ZA ^2 calculation}
	&\sigma_{Z^{\mathrm{A}}}^2 = \sum_{m=0}^{\imax-1}\frac{1}{\bar{M}}\myExp{\sigma_{z^{\mathrm{A}}}^2[m]} ,
\end{align}
where the cross-terms are suppressed due to 
the whiteness of $ z_n^{\mathrm{A}}[m] $, as illustrated in Lemma \ref{lm: sn[m] zn[m] uncorrelated}. As also shown in Lemma \ref{lm: sn[m] zn[m] uncorrelated}, $ \sigma_{z^{\mathrm{A}}}^2[m] $ is a discrete function of $ m $. 
Based on the uniform distribution of $  l_p  $ as stated in the condition of Lemma \ref{lm: variance of Snl ZAnl Zbnl Wnl}, $ \sigma_{z^{\mathrm{A}}}^2[m] $ takes each value in Table \ref{tab: value of variance of zAm} with the same probability $ \frac{1}{P} $. This further leads to 
\begin{align} \label{eq: sigma ZA m 2}
	 \myExp{\sigma_{z^{\mathrm{A}}}^2[m]}&=\frac{\sigma_d^2}{P} \left( \sigma_{P-1}^2 + \sum_{p=P-2}^{P-1}\sigma_p^2 + \cdots +\sum_{p=0}^{P-1}\sigma_p^2 \right)
	\nonumber\\
	&=\frac{\sigma_d^2}{P}\sum_{p=0}^{P-1}(p+1)\sigma_p^2.
\end{align}
Substituting (\ref{eq: sigma ZA m 2}) into (\ref{eq: sigma ZA ^2 calculation}), we obtain the variance of $ Z_n^{\mathrm{A}}[l] $, as given in (\ref{eq: sigma ZA ^2}). Similarly, $ \sigma_{Z^{\mathrm{B}}}^2 $ can be derived. The details are suppressed here.

\subsection{Proof of Lemma \ref{lm: determine k}}\label{app: proof of lemma determining k}
The probability $ \myProb{|\mk \tilde{S}_n[l]|\le 1} $ is equivalent to $ \myProb{|\mk \tilde{S}_n[l]|^2 \le 1 } $ which can be calculated as
\begin{align}
	& \frac{1}{\tilde{M}}=\myProb{|\mk \tilde{S}_n[l]|\le 1}=\myProb{|\mk \tilde{S}_n[l]|^2\le 1}\nonumber\\
	& =\myProb{\underbrace{\mk^2\frac{\sigma_d^2}{2}}_{1/z} \underbrace{\left( \frac{\Re\{\tilde{S}_n[l]\}^2}{\sigma_d^2/2} + \frac{\Im\{\tilde{S}_n[l] \}^2}{\sigma_d^2/2} \right)}_{x} \le 1} = \myProb{x\le z}\nonumber\\
	& = 1-e^{-\frac{1}{\mk^2\sigma_d^2}} , \nonumber
\end{align}
where $ \tilde{S}_n[l]\sim\mathcal{CN}(0,\sigma_d^2) $ is illustrated in Lemma \ref{lm: variance of Snl ZAnl Zbnl Wnl}, $ x $ then conforms to a chi-squared distribution with two degrees of freedom, and the last result is based on the fact that $ \myProb{x\le z}=1-e^{-z/2} $ \cite{book_simon2007probability}. Solving the above equation results in (\ref{eq: k the factor}).

\subsection{Proof of Proposition \ref{pp: variance of complex normal ratio}}\label{app: proof of proposition on variance of a complex gaussian ratio}

By definition, we have 
\begin{align} \label{eq: z=zr+jzi}
	z = \frac{x}{y}=\frac{xy*}{|y|}=\underbrace{\frac{x_{\mathrm{r}}y_{\mathrm{r}} + x_{\mathrm{i}}y_{\mathrm{i}}  }{|y|}}_{z_{\mathrm{r}}} - \mj\underbrace{\frac{(x_{\mathrm{i}}y_{\mathrm{r}}-x_{\mathrm{r}}y_{\mathrm{i}})}{|y|}}_{z_{\mathrm{i}}},
\end{align}
where adding the subscripts, $ (\cdot)_{\mathrm{r}} $ and $ (\cdot)_{\mathrm{i}} $, to a complex variable denotes the real and imaginary parts of the variable, respectively. The variance of $ z $, denoted by $ \sigma_z^2 $, can be calculated as $ \sigma_z^2=\sigma_{z_{\mathrm{r}}}^2 + \sigma_{z_{\mathrm{i}}}^2 $, where $ \sigma_{z_{\mathrm{r}}}^2 ~ (\mathrm{or~}\sigma_{z_{\mathrm{i}}}^2) $
denotes the variance of $ z_{\mathrm{r}}~(\mathrm{or~}z_{\mathrm{i}}) $. 
Since $ z_{\mathrm{r}} $ and $ z_{\mathrm{i}} $ have the same PDF \cite{ComplexGaussianRatio_2010Globecom} and occupy the same region, they have the same variance as well. Thus, we only illustrate the calculation of $ \sigma_{z_{\mathrm{r}}}^2 $ below.

With reference to \cite[Eq. (18)]{ComplexGaussianRatio_2010Globecom}, the cumulative density function of $ z_{\mathrm{r}} $ can be expressed as $ F_{z_{\mathrm{r}}}(u) = {(\lambda(u)+1)}/{2} $, 
where $ \lambda(u)=\frac{ u}{\sqrt{\rho +  u^2}} $ and $ \rho = \sigma_x^2/\sigma_y^2 $. Taking the derivative of $ F_{z_{\mathrm{r}}}(u) $ w.r.t. $ u $, we obtain the PDF of $ {z_{\mathrm{r}}} $, as given by
\begin{align}
	f_{z_{\mathrm{r}}}(u) = \frac{\rho}{2(\rho + u^2)^{\frac{3}{2}}}. \nonumber
\end{align}	
The PDF is an even function of $ u $ and hence $ \myExp{z_{\mathrm{r}}}=0 $. 
Directly calculating the variance of $ z_{\mathrm{r}} $ with an infinite region of $ u $ will results in an infinite variance. Thus, we put a finite upper limit on $ |z_{\mathrm{r}}| $ and approximate its variance as follows,
\begin{align}
	& \sigma_{z_{\mathrm{r}}}^2 = 2\int_{0}^{\bar{u}} u^2 f_{z_{\mathrm{r}}}(u)\mathrm{d}u = \rho^2  {\left(\mathrm{asinh}\left(\frac{\bar{u} }{\sqrt{\rho} }\right)-\frac{\bar{u} }{\sqrt{\rho +{\bar{u} }^2 }}\right)}  \nonumber\\
	& \myEqualOverset{\bar{u}=\bar{k}\sqrt{\rho}} 
	\rho  {\left(\mathrm{asinh}\left(\bar{k}\right)-\frac{\bar{k} }{\sqrt{1 +{\bar{k} }^2 }}\right)}
	 \myApprOverset{\bar{k}\gg 1} \rho \left(  \ln(2\bar{k}) - 1  \right), \nonumber
\end{align}
where, under $ \bar{k}\gg 1 $, two approximations are used, i.e., $ \mathrm{asinh}\left(\bar{k}\right)\approx \ln\left( 2\bar{k} \right) $ and $ \frac{\bar{k} }{\sqrt{1 +{\bar{k} }^2 }}\approx 1 $. 
Based on the PDF $ f_{z_{\mathrm{r}}}(u) $, the probability that $ |z_{\mathrm{r}}|\le \bar{u} $ can be calculated as 
\begin{align}
	\mP\{|z_{\mathrm{r}}|\le \bar{u}\} = \int_{-\bar{u}}^{\bar{u}} 	f_{z_{\mathrm{r}}}(u) \mathrm{d}u = \frac{\bar{u} }{\sqrt{{\bar{u} }^2 +\rho }} = \frac{\bar{k} }{\sqrt{{\bar{k} }^2 + 1 }}, \nonumber
\end{align}
where the substitution $  \bar{u}=\bar{k}\sqrt{\rho} $
is performed.
Assume that $ \mP\{|z_{\mathrm{r}}|\le \bar{u}\}=1-\epsilon $, where $ \epsilon $ is a sufficiently small probability claimed in the condition of Proposition \ref{pp: variance of complex normal ratio}. Then we can solve that $ \bar{k}=\sqrt{\frac{\bar{\epsilon}}{1-\bar{\epsilon}^2}} $ and, moreover, 
\begin{align}
	\sigma_{z_{\mathrm{r}}}^2 \approx \rho \left(  \ln\left( \sqrt{\frac{4\bar{\epsilon}}{1-\bar{\epsilon}^2}} \right) - 1  \right),~\mathrm{s.t.}~\bar{\epsilon}=1-\epsilon.
\end{align}
As mentioned earlier, $ z_{\mathrm{i}} $ has the same variance as $ z_{\mathrm{r}} $. Thus, the final variance of $ z $ becomes the one given in (\ref{eq: signa z^2 result}).

\subsection{Calculating $ \Pi $ given in (\ref{eq: sigma_S^2})} \label{app: calculating Pi}
Based on the expression of $ \Pi $ given in (\ref{eq: sigma_S^2}), we can have
\begin{align} \label{eq: Pi}
	& \Pi \myEqualOverset{(a)} \sum_{p=0}^{P-1}\sigma_p^2\myExp{ \sum_{\substack{m_1=0\\m_2=m_1}}^{\bar{M}-1}+ \sum_{\substack{m=|m_1-m_2|\\=1}}^{\bar{M}-1} \substack{(\bar{M}-m)\times \\
			(e^{\mj 2\pi\nu_{p} m T_{\mathrm{s}}} + e^{-\mj 2\pi\nu_{p} m T_{\mathrm{s}}})} } \nonumber\\
	& = \sum_{p=0}^{P-1}\sigma_p^2\left(\bar{M} + 2\sum_{\substack{m=1}}^{\bar{M}-1}(\bar{M}-m){\cos(2\pi \nu_p mT_{\mathrm{s}})}\right)\nonumber\\
	& \myApprOverset{\omega_p=2\pi\nu_pT_{\mathrm{s}}} 
	\sum_{p=0}^{P-1}\sigma_p^2
	\left(\bar{M} + 2\sum_{\substack{m=1}}^{\bar{M}-1}(\bar{M}-m){\left(1-\frac{m^2\omega_p^2}{2}\right)}\right)\nonumber\\
	& \myEqualOverset{(b)} \bar{M}^2\Big( \sigma_P^2 + \ma - \ma\bar{M}^2 \Big),
\end{align}
where $ \ma =\sum_{p=0}^{P-1}\sigma_p^2\frac{\omega_p^2}{12}=\frac{\pi^2 T_{\mathrm{s}}^2}{3}\sum_{p=0}^{P-1}\sigma_p^2 \nu_p^2$ and $\sigma_P^2=\sum_{p=0}^{P-1}\sigma_p^2  $. 
Note that the result $ \myEqualOverset{(a)} $ is obtained by separately calculating the cases of $ m_1=m_2 $ and $ m_1\ne m_2 $; the approximation is obtained based on the second-degree Taylor expansion of the cosine function w.r.t. $ m $; and 
the result $ \myEqualOverset{(b)} $
is obtained based on the following formula of summations: $ \sum_{m=1}^{\bar{M}-1} m = \bar{M}^2/2-\bar{M}/2 $, $ \sum_{m=1}^{\bar{M}-1} m^2 = \bar{M}^3/3 - \bar{M}^2/2 + \bar{M}/6 $ and $ \sum_{m=1}^{\bar{M}-1} m^3 = \bar{M}^4/4 - \bar{M}^3/2 + \bar{M}^2/4 $ \cite{book_rosen2017handbook_discreteMathCombinatorial}.

\subsection{Deriving $ \Gamma_1=\sigma_P^2 $ for (\ref{eq: Gamma})} \label{app: deriving Gamma_1=1}
For convenience, we provide below the expression of $ \Gamma_1 $, 
\begin{align}
	\sum_{p=0}^{P-1}\sigma_p^2{\left(   
	\begin{array}{l}
		\sum_{{l'=0}}^{\bar{M}-1}\sum_{m_1=0}^{\bar{M}-1} \sum_{m_2=0}^{\bar{M}-1} \\ [1mm]
		~~e^{\mj 2\pi\nu_{p} (m_1-m_2) T_{\mathrm{s}}} \frac{\myDFT{\bar{M}}{(l-l')(m_1-m_2)}}{\bar{M}^{\frac{3}{2}}}  
	\end{array}\right)}, \nonumber
\end{align}
where we notice that the summations enclosed in the round brackets can be calculated first. 
Similar to the way $ \Pi $ is calculated in Appendix \ref{app: calculating Pi}, we calculate $ \Gamma_1 $ by considering $ m_1=m_2 $ and $ m_1\ne m_2 $. For the first case, we have 
\[
\sum_{{l'=0}}^{\bar{M}-1}\sum_{\substack{m_1=0\\m_2=m_1}}^{\bar{M}-1}  e^{\mj 2\pi\nu_{p} (m_1-m_2) T_{\mathrm{s}}} \frac{\myDFT{\bar{M}}{(l-l')(m_1-m_2)}}{\bar{M}^{\frac{3}{2}}} =1,
\]
where we have used $ \myDFT{\bar{M}}{(l-l')\times 0}=\frac{1}{\sqrt{\bar{M}}} $; see (\ref{eq: tilde Sn[l]}).
For the case of $ m_1\ne m_2 $, the summation w.r.t. $ l' $ is always zero since the summands, given $ \forall m_1,m_2 $, are the samples of a complex exponential signal within an integer multiple of periods. Combining the two cases, we have $ \Gamma_1 = \sum_{p=0}^{P-1}\sigma_p^2\times 1=\sigma_P^2 $.

\subsection{Proving $ \myExp{\left(\frac{Z^{\mathrm{A}}_n[l]}{\mk \tilde{S}_n[l]}\right)^*\frac{Z^{\mathrm{B}}_n[l]}{\mk \tilde{S}_n[l]}}  \approx 0 $
 } \label{app: E{ZA/S ZB/S}=0}
For convenience, we define some shorthand expressions: $ Z^{\mathrm{A}}_n[l]=a+\mj b  $, $ Z^{\mathrm{B}}_n[l]=c+\mj d $ and $ \tilde{S}_n[l]=e+\mj f $. 
Based on (\ref{eq: breve ZnAl ZnBl Wnl}), we have 
\begin{align}
	\frac{Z^{\mathrm{A}}_n[l](Z^{\mathrm{B}}_n[l])^*}{\mk^2 |\tilde{S}_n[l]|^2} = \frac{\overbrace{(a c+b d)+\mj (bc-ad)}^{R}}{\underbrace{\mk^2(e^2+f^2)}_{S}}, \nonumber
\end{align}
The expectation of $ R/S $ can be calculated as \cite{Note_seltman2012approximations}
\begin{align}\label{eq: E()=mu/nu-sigma...}
	\myExp{R/S} \approx  \mu_R/\mu_S - \sigma_{RS}/\mu_S^2 + \sigma_S^2\mu_R/\mu_S^3,
\end{align}
where $ \mu_R $ and $ \mu_S $ are means, $ \sigma_{RS} $ denotes covariance and $ \sigma_S^2 $ denotes variance. 
Based on Lemma \ref{lm: sn[m] zn[m] uncorrelated}, 
$ a $, $ b $, $ c $, $ d $, $ e $ and $ f $ are independent centered Gaussian variables. 
Therefore, we have $ \mu_{R}=0 $ and moreover 
\begin{align}
	&\sigma_{RS} = \myExp{(R-\mu_R)^*(S-\mu_S)} = \myExp{R^*S} \nonumber\\
	&= \myExp{{\left({\left(a d-b c\right)} {\left(e^2 +f^2 \right)}\right)} \mj+{\left(a c+b d\right)} {\left(e^2 +f^2 \right)}}=0. \nonumber
\end{align}
These further lead to $ \myExp{\left(\frac{Z^{\mathrm{A}}_n[l]}{\mk \tilde{S}_n[l]}\right)^*\frac{Z^{\mathrm{B}}_n[l]}{\mk \tilde{S}_n[l]}}  \approx 0  $. Likewise, we can show $ \myExp{\left(\frac{Z^{\mathrm{B}}_n[l]}{\mk \tilde{S}_n[l]}\right)^*\frac{Z^{\mathrm{A}}_n[l]}{\mk \tilde{S}_n[l]}}\approx 0 $.

\subsection{Proof of Proposition \ref{pp: concave numerator in sinr}} \label{app: proof of proposition on concave numerator function}

Based on (\ref{eq: sigma_S^2}), we have 
\begin{align} \nonumber
	f(\bar{M}) 
	\approx \frac{ \bar{M}I  }{ (\bar{M}+Q) } \frac{1 }{ \mk^2 } \Big( \sigma_P^2+\ma - \ma \bar{M}^2 \Big)
\end{align}
where the approximation is due to $ \tilde{N} = \myRound{\frac{I}{\tilde{M}}}\approx\frac{I}{\tilde{M}}=\frac{I}{\bar{M}+Q} $. The first derivative of $ f(\bar{M}) $ w.r.t. $ \bar{M} $ is
\begin{align}
	f'(\bar{M}) 
	&=\frac{-\mC(2 {\bar{M} }^3  \ma + 3 Q {\bar{M} }^2  \ma - Q \ma - Q\sigma_P^2)}{{{\left(\bar{M} +Q\right)}}^2 },
\end{align}
and the second derivative is 
\begin{align}
	f''(\bar{M}) 
	& = -\frac{2\mC {\left({\bar{M} }^3  \ma +3 {\bar{M} }^2  Q \ma +3 \bar{M}  Q^2  \ma +Q \ma +Q\sigma_P^2\right)}}{{{\left(\bar{M} +Q\right)}}^3 }
\end{align}
where $ \mC(>0) $ absorbs the $ \bar{M} $-independent coefficients.
Since $ f''(\bar{M})<0 $ always holds, 
$ f'(\bar{M}) $ monotonically decreases. We notice that $ f'(0)>0 $. Thus, there exists $ \bar{M}_f^* $ such that $ f'(\bar{M})>0 $ for $ \bar{M}<\bar{M}_f^* $, $ f'(\bar{M})=0 $ for $ \bar{M}=\bar{M}_f^* $, and $ f'(\bar{M})<0 $ for $ \bar{M}>\bar{M}_f^* $. The value of $ \bar{M}_f^* $ can be determined by solving the equation $ f'(\bar{M})=0 $ which is essentially 
\[2 {\bar{M} }^3  \ma + 3 Q {\bar{M} }^2  \ma - Q \ma - Q\sigma_P^2  = 0.\]
Directly using the cubic formula leads to a solution with a complex structure and does not provide much insight. 
To this end, we consider the case of $ \bar{M}\gg 3Q/2 $ and simplify the equation by dropping the quadratic term, obtaining $ 2 {\bar{M} }^3  a  -Q a -Q\sigma_P^2  = 0 $. Thus, an approximation of $ \bar{M}_f^* $ is achieved as 
\begin{align} \nonumber
	\bar{M}_f^* \myApprOverset{\bar{M}\gg 3Q/2} \sqrt[3]{\frac{Q\ma + Q\sigma_P^2}{2\ma}}.
\end{align}

\subsection{Proof of Proposition \ref{pp: convex denominator in sinr}} \label{app: proof of proposition on convex denominator in sinr}

Based on (\ref{eq: sigma_I^2}), (\ref{eq: sigma_Z^2}) and (\ref{eq: sigma_W^2}), we have 
\begin{align}
	&g(\bar{M}) =\frac{\mb(\epsilon)}{\mk^2} \left(  (\ma\bar{M}^2 - \ma) + \frac{Q(P+1)\sigma_P^2}{\bar{M}P } + \frac{\sigma_w^2}{\sigma_d^2}\right)
	\nonumber\\
	& = \frac{\mb(\epsilon)}{\mk^2}\frac{{\ma} {\bar{M} }^3 +{\left(\frac{\sigma_w^2}{\sigma_d^2}-\ma  \right)} \bar{M} +\frac{Q(P+1)\sigma_P^2}{P }}{\bar{M} }. \nonumber
\end{align}
The first derivative of $ g(\bar{M}) $ is 
\begin{align}
	g'(\bar{M})=  \frac{\frac{\mb(\epsilon)}{\mk^2} \left(
		2{\bar{M} }^3  \ma -\frac{Q(P+1)\sigma_P^2}{P } 
		\right) 
	}{{\bar{M} }^2 }. \nonumber
\end{align}
Moreover, we can readily validate that the second derivative of $ g(\bar{M}) $ is constantly positive, which is suppressed here for brevity. 
Therefore, there exists $ \bar{M}_g^* $ such that $ g'(\bar{M})<0 $ for $ \bar{M}<\bar{M}_g^* $, $ g'(\bar{M})=0 $ for $ \bar{M}=\bar{M}_g^* $, and $ g'(\bar{M})>0 $ for $ \bar{M}>\bar{M}_g^* $. Solving $ g'(\bar{M})=0 $, we obtain 
\begin{align} \nonumber
	\bar{M}_g^* = \sqrt[3]{{Q(P+1)\sigma_P^2}\Big/{(2\ma P)}}.
\end{align}

\textbb{
\subsection{Approximate CRLBs for the estimates obtained in (\ref{eq: hat l_p hat nu_p})} \label{app: crlb}

From Section \ref{sec: sensing framework parameter estimation}, we see that the velocity and range estimations are both turned into frequency estimations. Thus, we can use the CRLB of a single-tone frequency estimator to derive the CRLBs of $ \hat{\nu}_p $ and $ \hat{l}_p $ obtained in (\ref{eq: hat l_p hat nu_p}).
As the derivations for the two estimates are very similar, we only take $ \hat{\nu}_p = (\tilde{n}_p + \hat{\epsilon}_p)\Big/(\tilde{N}\tilde{M}T_{\mathrm{s}})$ for an illustration. 
According to \cite{Kai_freqEst2020CL}, the CRLB of $ (\tilde{n}_p + \hat{\epsilon}_p) $ can be given by $ 6/(4\pi^2 \tilde{\gamma}) $, where $ \tilde{\gamma} $ is the SNR of $ \breve{X}_{\tilde{n}_p^{\pm}}^{(p)}[\tilde{l}_{p}] $ obtained in (\ref{eq: breve X _{tilde n_p +-} [tilde l_p]}).
The SNR $ \tilde{\gamma} $ can be approximated by suppressing $ \sigma_{\mathrm{I}}^2 $ and $ \sigma_{\mathrm{Z}}^2 $ in (\ref{eq: gamma overall SINR}), leading to $ \tilde{\gamma} = \frac{\bar{M}\tilde{N}\sigma_{\mathrm{S}}^2}{\sigma_{{W}}^2} $, where $ \sigma_{\mathrm{S}}^2 $ is given in (\ref{eq: sigma_S^2}) and $ \sigma_W^2 $
in (\ref{eq: sigma_W^2}). 
Given $ \hat{\nu}_p = (\tilde{n}_p + \hat{\epsilon}_p)\Big/(\tilde{N}\tilde{M}T_{\mathrm{s}}) $, we have $ \myCRLB{ \hat{\nu}_p } = \myCRLB{\tilde{n}_p + \hat{\epsilon}_p}/(\tilde{N}\tilde{M}T_{\mathrm{s}})^2 $. Since $ \nu_p=2v_p/\lambda $, we further have $ \myCRLB{\hat{v}_p}=\frac{\lambda^2}{4} \myCRLB{\hat{\nu}_p} $.
Combining the above analyses gives the CRLB in (\ref{eq: crlb velocity estimation}).
}

\bibliographystyle{IEEEtran}
\bibliography{IEEEabrv,./bib_JCAS.bib}

% Generated by IEEEtran.bst, version: 1.14 (2015/08/26)
\begin{thebibliography}{10}
\providecommand{\url}[1]{#1}
\csname url@samestyle\endcsname
\providecommand{\newblock}{\relax}
\providecommand{\bibinfo}[2]{#2}
\providecommand{\BIBentrySTDinterwordspacing}{\spaceskip=0pt\relax}
\providecommand{\BIBentryALTinterwordstretchfactor}{4}
\providecommand{\BIBentryALTinterwordspacing}{\spaceskip=\fontdimen2\font plus
\BIBentryALTinterwordstretchfactor\fontdimen3\font minus
  \fontdimen4\font\relax}
\providecommand{\BIBforeignlanguage}[2]{{%
\expandafter\ifx\csname l@#1\endcsname\relax
\typeout{** WARNING: IEEEtran.bst: No hyphenation pattern has been}%
\typeout{** loaded for the language `#1'. Using the pattern for}%
\typeout{** the default language instead.}%
\else
\language=\csname l@#1\endcsname
\fi
#2}}
\providecommand{\BIBdecl}{\relax}
\BIBdecl

\bibitem{chapter_sari2020industrialnowFutureTrends}
A.~Sari, A.~Lekidis, and I.~Butun, ``Industrial networks and {IIoT}: Now and
  future trends,'' in \emph{Industrial IoT}.\hskip 1em plus 0.5em minus
  0.4em\relax Springer, 2020, pp. 3--55.

\bibitem{chapter_IIoT_celebi2020wireless}
H.~B. Celebi, A.~Pitarokoilis, and M.~Skoglund, ``Wireless communication for
  the industrial {IoT},'' in \emph{Industrial IoT}.\hskip 1em plus 0.5em minus
  0.4em\relax Springer, 2020, pp. 57--94.

\bibitem{IIOT_industrialWirelessNetworkSurvey}
M.~Raza, N.~Aslam, H.~Le-Minh, S.~Hussain, Y.~Cao, and N.~M. Khan, ``A critical
  analysis of research potential, challenges, and future directives in
  industrial wireless sensor networks,'' \emph{IEEE Commun. Surveys Tutor.},
  vol.~20, no.~1, pp. 39--95, 2017.

\bibitem{IIoT_wirelessNetworkDesign}
Y.~Liu, M.~Kashef, K.~B. Lee, L.~Benmohamed, and R.~Candell, ``Wireless network
  design for emerging {IIoT} applications: Reference framework and use cases,''
  \emph{Proc. IEEE}, vol. 107, no.~6, pp. 1166--1192, 2019.

\bibitem{Kai_overviewFHMIMO_DFRC2020AES}
K.~Wu, J.~A. Zhang, X.~Huang, and Y.~J. Guo, ``Frequency-hopping mimo
  radar-based communications: An overview,'' \emph{arXiv preprint
  arXiv:2006.07559}, 2020.

\bibitem{FanLiu_overview2020TCOM}
F.~{Liu}, C.~{Masouros}, A.~P. {Petropulu}, H.~{Griffiths}, and L.~{Hanzo},
  ``Joint radar and communication design: Applications, state-of-the-art, and
  the road ahead,'' \emph{IEEE Trans. Commun.}, vol.~68, no.~6, pp. 3834--3862,
  2020.

\bibitem{Kai_rahman2020enablingSurvey}
M.~L. Rahman, J.~A. Zhang, K.~Wu, X.~Huang, Y.~J. Guo, S.~Chen, and J.~Yuan,
  ``Enabling joint communication and radio sensing in mobile networks--a
  survey,'' \emph{arXiv preprint arXiv:2006.07559}, 2020.

\bibitem{6G_XiaohuYou2021towards}
X.~You, C.-X. Wang, J.~Huang, X.~Gao, Z.~Zhang, M.~Wang, Y.~Huang, C.~Zhang,
  Y.~Jiang, J.~Wang \emph{et~al.}, ``Towards 6g wireless communication
  networks: Vision, enabling technologies, and new paradigm shifts,''
  \emph{Science China Information Sciences}, vol.~64, no.~1, pp. 1--74, 2021.

\bibitem{OTFS_Magazine_wei2020orthogonal}
Z.~Wei, W.~Yuan, S.~Li, J.~Yuan, G.~Bharatula, R.~Hadani, and L.~Hanzo,
  ``Orthogonal time-frequency space modulation: A full-diversity next
  generation waveform,'' \emph{arXiv preprint arXiv:2010.03344}, 2020.

\bibitem{JCAS_akan2020internetOfRadar_iot}
O.~B. Akan and M.~Arik, ``Internet of radars: Sensing versus sending with joint
  radar-communications,'' \emph{IEEE Communications Magazine}, vol.~58, no.~9,
  pp. 13--19, 2020.

\bibitem{JCAS_cui2021integratingJCAS4IoT}
Y.~Cui, F.~Liu, X.~Jing, and J.~Mu, ``Integrating sensing and communications
  for ubiquitous iot: Applications, trends and challenges,'' \emph{arXiv
  preprint arXiv:2104.11457}, 2021.

\bibitem{IIoT_droneskumar2021internet}
A.~Kumar and P.~L. Mehta, ``Internet of drones: An engaging platform for
  iiot-oriented airborne sensors,'' in \emph{Smart Sensors for Industrial
  Internet of Things}.\hskip 1em plus 0.5em minus 0.4em\relax Springer, 2021,
  pp. 249--270.

\bibitem{OTFS_keskin2021radarTimeDomainICIisi}
M.~F. Keskin, H.~Wymeersch, and A.~Alvarado, ``Radar sensing with otfs:
  Embracing isi and ici to surpass the ambiguity barrier,'' \emph{arXiv
  preprint arXiv:2103.16162}, 2021.

\bibitem{OTFS_hadani2018otfs_book_chapter}
R.~Hadani and A.~Monk, ``Otfs: A new generation of modulation addressing the
  challenges of 5g,'' \emph{arXiv preprint arXiv:1802.02623}, 2018.

\bibitem{OTFS_jcas2020twc}
L.~Gaudio, M.~Kobayashi, G.~Caire, and G.~Colavolpe, ``On the effectiveness of
  {OTFS} for joint radar parameter estimation and communication,'' \emph{IEEE
  Trans. Wireless Commun.}, vol.~19, no.~9, pp. 5951--5965, 2020.

\bibitem{OTFS_yuan2021integratedSensingCOmmunicationOTFS}
W.~Yuan, Z.~Wei, S.~Li, J.~Yuan, and D.~W.~K. Ng, ``Integrated sensing and
  communication-assisted orthogonal time frequency space transmission for
  vehicular networks,'' \emph{arXiv preprint arXiv:2105.03125}, 2021.

\bibitem{OTFS_Raviteja2019TVT_embeddedPilotChannelEstimation}
P.~{Raviteja}, K.~T. {Phan}, and Y.~{Hong}, ``Embedded pilot-aided channel
  estimation for {OTFS} in delay–doppler channels,'' \emph{IEEE Trans. Veh.
  Techn.}, vol.~68, no.~5, pp. 4906--4917, 2019.

\bibitem{DFRC_dsss2011procIeee}
C.~{Sturm} and W.~{Wiesbeck}, ``Waveform design and signal processing aspects
  for fusion of wireless communications and radar sensing,'' \emph{Proc. IEEE},
  vol.~99, no.~7, pp. 1236--1259, 2011.

\bibitem{Kai_ofdmSensingSPM}
K.~Wu, J.~A. Zhang, X.~Huang, and Y.~J. Guo, ``A low-complexity method for
  {FFT}-based {OFDM} sensing,'' \emph{arXiv preprint arXiv:2105.13596}, 2021.

\bibitem{OTFS_Raviteja2019TVTpulseShapingRCP}
P.~Raviteja, Y.~Hong, E.~Viterbo, and E.~Biglieri, ``Practical pulse-shaping
  waveforms for reduced-cyclic-prefix otfs,'' \emph{IEEE Trans. Veh. Techn.},
  vol.~68, no.~1, pp. 957--961, 2019.

\bibitem{DFRC_SC_OFDM}
Y.~Zeng, Y.~Ma, and S.~Sun, ``Joint radar-communication with cyclic prefixed
  single carrier waveforms,'' \emph{IEEE Trans. Veh. Techn.}, vol.~69, no.~4,
  pp. 4069--4079, 2020.

\bibitem{book_richards2010principlesModernRadar}
M.~A. Richards, J.~Scheer, W.~A. Holm, and W.~L. Melvin, \emph{Principles of
  modern radar}.\hskip 1em plus 0.5em minus 0.4em\relax Citeseer, 2010.

\bibitem{Gaussian_OFDMenvelope}
S.~{Wei}, D.~L. {Goeckel}, and P.~A. {Kelly}, ``Convergence of the complex
  envelope of bandlimited {OFDM} signals,'' \emph{IEEE Trans. Information
  Theory}, vol.~56, no.~10, pp. 4893--4904, 2010.

\bibitem{book_oppenheim1999discrete}
A.~V. Oppenheim, \emph{Discrete-time signal processing}.\hskip 1em plus 0.5em
  minus 0.4em\relax Pearson Education India, 1999.

\bibitem{Kai_freqEst2020CL}
K.~{Wu}, W.~{Ni}, J.~A. {Zhang}, R.~P. {Liu}, and Y.~J. {Guo}, ``Refinement of
  optimal interpolation factor for {DFT} interpolated frequency estimator,''
  \emph{IEEE Commun. Lett.}, pp. 1--1, 2020.

\bibitem{Kai_padeFreqEst2021TVT}
K.~Wu, J.~A. Zhang, X.~Huang, and Y.~J. Guo, ``Accurate frequency estimation
  with fewer {DFT} interpolations based on pad\'e approximation,'' \emph{arXiv
  preprint arXiv:2105.13567}, 2021.

\bibitem{book_ahmadi2019_5G}
S.~Ahmadi, \emph{{5G} {NR}: Architecture, Technology, Implementation, and
  Operation of {3GPP} New Radio Standards}.\hskip 1em plus 0.5em minus
  0.4em\relax Academic Press, 2019.

\bibitem{Kai_integrateSensingIntoCom2021JSAC}
K.~Wu, J.~A. Zhang, X.~Huang, and Y.~J. Guo, ``Integrating low-complexity and
  flexible sensing into communication systems,'' \emph{arXiv preprint
  arXiv:2109.04109}, 2021.

\bibitem{FreqEst_AMexMultipleTones2017SP}
S.~Ye and E.~Aboutanios, ``Rapid accurate frequency estimation of multiple
  resolved exponentials in noise,'' \emph{Signal Process.}, vol. 132, pp.
  29--39, 2017.

\bibitem{FreqEst_multiToneQSE2020}
A.~{Serbes} and K.~{Qaraqe}, ``A fast method for estimating frequencies of
  multiple sinusoidals,'' \emph{IEEE Signal Process. Lett.}, vol.~27, pp.
  386--390, 2020.

\bibitem{book_van2004optimum}
H.~L. Van~Trees, \emph{Optimum array processing: Part IV of detection,
  estimation, and modulation theory}.\hskip 1em plus 0.5em minus 0.4em\relax
  John Wiley \& Sons, 2004.

\bibitem{book_simon2007probability}
M.~K. Simon, \emph{Probability distributions involving Gaussian random
  variables: A handbook for engineers and scientists}.\hskip 1em plus 0.5em
  minus 0.4em\relax Springer Science \& Business Media, 2007.

\bibitem{liu2019robust}
Y.~Liu, G.~Liao, and Z.~Yang, ``Robust {OFDM} integrated radar and
  communications waveform design based on information theory,'' \emph{Signal
  Processing}, vol. 162, pp. 317--329, 2019.

\bibitem{liu2017multiobjective}
Y.~Liu, G.~Liao, Z.~Yang, and J.~Xu, ``Multiobjective optimal waveform design
  for {OFDM} integrated radar and communication systems,'' \emph{Signal
  Processing}, vol. 141, pp. 331--342, 2017.

\bibitem{liu2017adaptive}
Y.~Liu, G.~Liao, J.~Xu, Z.~Yang, and Y.~Zhang, ``Adaptive {OFDM} integrated
  radar and communications waveform design based on information theory,''
  \emph{IEEE Communications Letters}, vol.~21, no.~10, pp. 2174--2177, 2017.

\bibitem{liu2020super}
Y.~Liu, G.~Liao, Y.~Chen, J.~Xu, and Y.~Yin, ``Super-resolution range and
  velocity estimations with ofdm integrated radar and communications
  waveform,'' \emph{IEEE Transactions on Vehicular Technology}, vol.~69,
  no.~10, pp. 11\,659--11\,672, 2020.

\bibitem{liu2019joint}
Y.~Liu, G.~Liao, Z.~Yang, and J.~Xu, ``Joint range and angle estimation for an
  integrated system combining mimo radar with ofdm communication,''
  \emph{Multidimensional Systems and Signal Processing}, vol.~30, no.~2, pp.
  661--687, 2019.

\bibitem{ComplexGaussianRatio_2010Globecom}
R.~J. {Baxley}, B.~T. {Walkenhorst}, and G.~{Acosta-Marum}, ``Complex gaussian
  ratio distribution with applications for error rate calculation in fading
  channels with imperfect csi,'' in \emph{2010 IEEE Global Telecommunications
  Conference GLOBECOM 2010}, 2010, pp. 1--5.

\bibitem{book_rosen2017handbook_discreteMathCombinatorial}
K.~H. Rosen, \emph{Handbook of discrete and combinatorial mathematics}.\hskip
  1em plus 0.5em minus 0.4em\relax CRC press, 2017.

\bibitem{Note_seltman2012approximations}
H.~Seltman, ``Approximations for mean and variance of a ratio,''
  \emph{unpublished note}, 2012.

\end{thebibliography}
\end{document}